# Even-odd alternative dispersions and beyond. Part II. Noninertial and inertial particles, and, astrophysical chirality analogy

Jian-Zhou Zhu (朱建州)

*[a]Su-Cheng Centre for Fundamental and Interdisciplinary Sciences Gaochun Nanjing 211316 China*

## Abstract

Particle transports in carriers with even-odd alternating dispersions (introduced in Part I) are investigated. For the third-order dispersion as in Korteweg-de-Vries (KdV), such alternating dispersion has the effects of not only regularizing the velocity from forming shock singularity (thus the attenuation of particle clustering strength) but also symmetrizing the oscillations (thus the corresponding skewness of the particle densities), among others, as demonstrated numerically. The analogy of such dispersion effects and consequences (on particle transports in particular) with those of helicity in Burgers turbulence, addressed in the context of astrophysics and cosmology, is made for illumination and promoting models. Both dispersion and helicity regularize the respective systems, and both are shown to be transferred by the drag to the flows of the respective inertial particles carried by the latter and to similarly affect the particle clustering. Among many details, a reward from studying particle transports is the understanding of the (asymptotic) $k^0$-scaling (equipartition among the wavenumbers, $k$s), before large-$k$ exponential decay, of the power spectrum of KdV solitons [resulting in the more general statement (valid beyond the KdV soliton and Burgers shock) that "a (one-dimensional) soliton is the derivative of a classical shock, just like the Dirac delta is the derivative of a step function"], motivated by the explanation of the the same scaling law of the particle densities as the apparent approximation of the Dirac deltas; while, the "shocliton" from the even-odd alternating dispersion in aKdV appears to be, indeed, *shock* ⊕ *soliton*, accordingly the decomposition of the averaged odd-mode spectrum, from sinusoidal initial field, into a $k^{-2}$ part for the shock and a $k^0$-scaling part for the solitonic pulses, only the latter being contained in the averaged even-mode spectrum.

## 1. Introduction

### 1.1. Backgrounds

We have proposed the dispersions with opposite signs for alternative Fourier components in the former communication [1] (hereafter "I"). Such "alternating dispersions" lead to novel features of the dynamics, such as the oscillations similar (with close amplitudes and frequencies/wavelengths) on both sides of a (dispersive) shock and the emergence of apparent shock and anti-shock duo, which may well model some real-world phenomena, including the quantum and plasma shocks. The new dispersive shock ("shocliton") is long-living, and the duo drift simultaneously, slowly (much slower than any of the soliton observed). A lot more studies are needed for deeper understanding and useful (if indeed) applications. For example, is the new shocliton fundamentally different to the classical (Burgers) shock? If yes, how? And, for the Korteweg-de Vries (KdV) model and that modified with the alternating dispersion (aKdV) as examples, are the KdV pulses fundamentally different to the aKdV oscillations, in terms of multi-scale spectra, say?

Before the other even more academic aspects, such as the issues of integrability and thermalization (related to the Fermi-Pasta-Ulam problem) remarked in I, we present here the exploration on the direction oriented towards applications, i.e., particle transports associated to astrophysical and cosmological chirality

---

and structure formation, which will also provide relevant fundamental insights and motivations on the study of the dispersive modles themselves.

Our purposes are two-fold: On the one hand, we would like to investigate the particle dynamics in flows characterized by such dispersion models, which is not only helpful in demonstrating and understanding of the properties of the alternating dispersion itself but also relevant to natural phenomena associated to particles in dispersive mediums; on the other hand, we hope to obtain insights for understanding and even modeling other physical processes with similar or analogous effects, such as some chiral effects on, say, the cosmic dust in astrophysics.

There are several facets in the backgrounds related to the first purpose. For the dispersive mediums and their mathematical model results, and, particularly the alternating dispersions, it may be sufficient to consult the discussions [mainly those related to the Korteweg-de-Vries (KdV) model which will also be used in this study] in I where readers can find many literatures in the bibliography there (with emphasis on the solitary waves/solitons though). Closely relevant literatures and results will be introduced in the next section (2) for a more technical discussion of the theoretical motivations and considerations. For an introductory discussion here, it is necessary to point out two relevant aspects of the dispersive effects: one is the regularization of the shock singularity that otherwise can present in the inviscid Burgers(-Hopf) solution, in a way different to the diffusion or adhesion in the context of cosmological structure formation (dispersive and dissipative regularizations are not always easily distinguishable in a physical system limited by the techniques [2]); the other is the formation of coherent structures (solitons) different to the dissipative shock, both may be analytically obtained with distinct mathematical techniques. When both of these regularizations present, it is natural to expect the emergence of classical dissipative shocks accompanied with oscillations, which appear to be mostly clearly represented by the well-known KdV-Burgers (KdVB) model. For particles carried by flows, it is a large topic connected with various research fields such as hydrodynamic and hydraulic engineering (e.g., sedimentation), ship and ocean engineering (e.g., bubbles), environments and atmosphere (e.g., sand storms, chemicals in and dust from combustions, aerosol and rain droplets), and, astrophysics and cosmology (e.g., dust in molecular clouds). Even with the models of passive transports neglecting the back reactions onto the flows, the mathematical analyses of such problems, especially when the flow is turbulent, are highly nontrivial, although remarkable analytical and physically illuminating results on the Kraichnan model [3], the problem of "passive scalars" (including density, tracer and vector) in a "rapidly" fluctuating (delta-correlated in time) Gaussian velocity field, have been obtained (c.f., e.g., Ref. [4] for a comprehensive review). There are still interesting theoretical progresses in the latter category (e.g., Ref. [5] on multifractal clustering in compressible flows relevant to our topic), especially on models extending that of Kraichnan to be more realistic, in this century, but, for particle transportation concerned here, the tools and results of a majority of the works have been based on or heavily relied on computer simulations: Such studies are overwhelming, with numerous publications from different disciplines which we cannot thoroughly review (again, closely relevant literatures and results will however be introduced a bit later). Rather, it appears efficient and sufficient for us to focus on the introduction of relevant studies by combining it to that of our second purpose, i.e., the chiral effects of particle-laden astrophysical flows or cosmological structure evolution.

For the chirality in astrophysics associated to our second purpose, with the helicity in the hydrodynamic level being an important aspect, it can be traced to the very fundamental (microscopic) laws of Nature, to the very early stage of the Universe, and can present in local regions or the whole of the latter [6] (see also Ref. [7] and references therein). We will be interested in the consequence of the compressibility reduction effect of helicity (or helicity "fastening" effect [8]), which, roughly speaking, refers to the less fraction of the compressive modes in the energy partition in a turbulence with more helicity (in the sense of absolute value — see below). Since our idea on the chiral effects of the inertial and noninertial particles carried by the neutral fluid can be naturally extended to the case of plasmas, our results will mainly be about the kinetic helicity, with brief remarks on the magnetic case.

The above remarks indicate that the two purposes are unified in some sense.

We consider the inertial and noninertial particles in the turbulent carrier which can be quite general, say, the compressible and incompressible neutral and ionized fluids, and, even quantum fluids and pressureless (magneto-)Burgers turbulence ("Burgulence" [9]) that in principle can be infinitely compressible [10, 11]. To isolate the helicity effect, we exclude the back reaction of the particles onto the carrier. That is, we limit ourselves to the passive transport issue. We focus on the neutral fluid case, but for comparison with the density scalar of the particles, we also present some results of tracer scalars (which can be a concentration).

We demonstrate that, since the passive density for inertial particles (with the backreaction or various couplings [12, 13, 14] neglected) is driven by some function(al) of the carrier flow velocity [15, 16, 17], the helicity of the carrier can also be transferred into the inertial particle flow (in the two-fluid or Eulerian formulation) which is then affected by the helicity, particularly the degree of compressibility.

In astrophysics, it is mere cliché to say the importance of particle transport problem for the formation of planets, stars and even larger cosmological structures, and we may quote that "dust contains a large fraction of the metals in the Universe, and is prominent in the interstellar medium...protoplanetary discs...and our Solar system... Dust physics is also key to understanding extinction and reddening in radiative transfer, feedback, and winds for both star formation and active galactic nuclei, galactic-chemistry, stellar evolution, interstellar heating and cooling, and more" [18], as represented also in various other studies, particularly intensively recently [19, 20, 21, 22, 23, 24, 25, 26, 27, 28, 29, 30, 31, 32]. We are oriented towards such issues where helicity and different dances of waves are important players.

As said, concerning the helicity fastening effect, various carriers can be considered, but to be definite and for simplicity, we use the flow governed by the Burgers equation. Using the Burgers carrier unifies the analysis of the flows by the fact that the fluid velocity of the transported inertial particles also satisfies the Burgers equation (up to a linear dragging term) in a reasonable model.

Actually, numerical tests of the basic idea of helicity fastening effect with the Burgers equation had been performed [33] (see Fig. 12 below), showing the reduction of the spectral ratio $E_{\parallel}/E_{kin}$ of the compressive modes as a measure of the compressibility. Consistent results were then obtained from another computation with different methods [34], extending to the magneto-Burgulence case with more systematic tests and analyses meant for a different theme. More complete understanding of the mechanism and applications of the helicity fastening effect are still underway. Here, we present a different and complementary branch of approaches with the idea of dispersion regularization and with the emphasis on the consequences on particle transports and astrophysical implications. The basic idea is the analogy between the dispersive regularization of the shocks in KdV(B) and the compressibility reduction (and thus regularization of the solution to some degree) with helicity in three-dimensional Burgers flows, and thus similarly the respective consequence on the particle transport.

The following sections are subsequently: Sec. 2 presents the theoretical motivations and considerations, with respectively Sec. 2.1 for the introduction of basic models, Sec. 2.1.1 for the discussions on the helicity fastening effect in Burgulence and its consequence on the transports of passive tracer and density scalars, Sec. 2.1.2 for the consequence on inertial particle clustering, Sec. 2.2 for the astrophysical applications and Sec. 2.3 for the analogy with the dispersive regularization effects; Sec. 3 offers numerical results on particles in KdV(B) and aKdV(B) systems, with the emphasis on the dispersive regularization effects on the clustering of the dust; Sec. 4 is a reward from studying the particle transports, with a better understanding of the multi-scale spectra of the dispersive oscillations and shocliton; and, finally, Sec. 5 is for further discussions, including in particular more on astrophysical chirality echoing and continuing the remarks in Sec. 2.

## 2. Particle transports connecting the chirality and dispersion issues

In the following subsections of this Sec. 2, we discuss our theoretical motivations and considerations, starting from the helicity fastening effect in three-dimensional Burgulence and its consequences on particle dynamics and ending with the dispersive effect in one-dimensional KdV(B), before turning to the next Sec. 3 for analysis with the numerical results of particles in aKdV(B).

*2.1. Consequences in particle transportation from the helicity fastening effect in 3D Burgulence*

The Burgers equation reads

$$\partial_t \boldsymbol{u} + \boldsymbol{u} \cdot \nabla \boldsymbol{u} = \nu \nabla^2 \boldsymbol{u} + \boldsymbol{f}, \tag{2.1}$$

where the forcing $\boldsymbol{f}$ can be used to control the injection of helicity $\mathcal{H} = (2V)^{-1} \iiint_V \nabla \times \boldsymbol{u} \cdot \boldsymbol{u} \, dV$ in the domain of volume $V$: the flow is helical if $\mathcal{H} \neq 0$, otherwise nonhelical; and larger $|\mathcal{H}|$ means more helical. With appropriate normalization, the kinetic viscosity coefficient $\nu$ would be simply the inverse of the Reynolds number. When the continuity equation for the mass density $\rho$ is included (see below), different invariance laws involving the density $\rho$ and velocity $\boldsymbol{u}$ can be derived [35]. Taking the curl of Eq. (2.1), we see that vorticity cannot be generated, if $\boldsymbol{f}$ is non-vortical.

To discuss the consequences of the helicity fastening effect on the transports of passive fields, we first consider two kinds of scalars, i.e., a tracer and a density. A density $\theta$ (for non-inertial particles, say) satisfies

$$(\partial_t - \kappa_\theta \nabla^2)\theta - \varphi_\theta = -\boldsymbol{u} \cdot \nabla \theta - \theta \nabla \cdot \boldsymbol{u} = -\nabla \cdot (\boldsymbol{u}\theta), \tag{2.2}$$

with $\varphi_\theta$ being the pumping and $\kappa_\theta$ the corresponding diffusion coefficient, while a passive tracer $c$ (as a concentration, i.e., mass ratio for the composition of the mixture of fluids with total density $\rho$, say) satisfies

$$(\partial_t - \kappa_c \nabla^2)c - \varphi_c = -\boldsymbol{u} \cdot \nabla c = -\nabla \cdot (c\boldsymbol{u}) + c\nabla \cdot \boldsymbol{u}, \tag{2.3}$$

with $\varphi_c$ being the corresponding pumping and $\kappa_c$ the diffusive coefficient. These two different kinds of scalars present distinct mathematical properties in weakly and strongly compressible regimes of $\boldsymbol{u}$ [36], but they are also closely related in some situations: while entropy and temperature behave as a tracer under suitable conditions, their gradients in one-dimension space are governed by the dynamics of a density transport [37], and $c\rho$ of a component of the mixture behaves as a density (see below). For the compressible Kraichnan model, phase transition of the cascade directions (directly towards smaller scales and inversely towards larger scales) of the tracer energy with the critical value of an appropriately defined measure for the degree of compressibility depending on the dimensionality can be predicted [38]. The passive scalars and vectors are different to the active ones with back reactions, and some two-dimensional cases have been compared face on face [39].

Note that with the continuity equation for the invariance of fluid mass with density $\rho$,

$$\partial_t \rho + \nabla \cdot (\rho \boldsymbol{u}) = 0 \tag{2.4}$$

which is indeed a passive density scalar in Burgers flows and with which, Eq. (2.3), with the diffusion and pumping neglected here for brevity, can be re-written in a conservative form with $\rho c$ as the density (say, of the composition of the mixture of fluids), i.e.,

$$\partial_t(\rho c) + \nabla \cdot (\rho c \boldsymbol{u}) = 0 \tag{2.5}$$

which is adopted by usual numerical literatures (e.g., Refs. [40, 41, 42]). [Our objectives are the viscous and diffusive cases, with presumably no real singularities.] In the physical context of positive-value scalars, the formulation in terms of logarithmic variables, such as $\ln c$ and $\ln \theta$, can also be used for both numerical and theoretical purposes (see, e.g., Ref. [43] and references therein, and, Sec. 3 below).

We also caution that some authors (e.g., Refs. [44, 45]) claimed a "tracer", probably meant for non-inertial/tracer particles, but actually work with a density scalar, which might cause confusion: indeed, when speaking about "tracer", we should distinguish the particles themselves and the physical variables quantifying some property of the particles. Note that for inertial particles, the above carrier $\boldsymbol{u}$ should be replaced by the inertial particle velocity field $\boldsymbol{v}$ (e.g., Refs. [12, 16] for particular discussions on the Eulerian and Lagrangian formulations), which we will come back to in Sec 2.1.2.

### 2.1.1. Consequences of the helicity fastening effect on the transports of noninertial particles

When the diffusion and pumping are neglected for the density and tracer scalars, i.e.,

$$\partial_t \theta + \nabla \cdot (\theta \boldsymbol{u}) = 0, \tag{2.6}$$

$$\partial_t c + \boldsymbol{u} \cdot \nabla c = 0, \tag{2.7}$$

they are straightforwardly solved in terms of smooth Lagrangian flows. [The $\theta$ equation in such ideal situation is just that of $\rho$, which is not surprising because, up to diffusion (neglected now), the noninertial particle simple trace the fluid particle. Note however that $\rho$ in general flows is not passive.] We can introduce the Lagrangian flow map (particle trajectory) $\boldsymbol{x}(\boldsymbol{\zeta}, t) = \boldsymbol{X}(t; \boldsymbol{\zeta}, 0)$ at time $t$, passing through the location $\boldsymbol{\zeta}$ at instant 0; that is, $\boldsymbol{x}(\boldsymbol{\zeta}, 0) = \boldsymbol{\zeta}$. We have $d\boldsymbol{x}/dt = \boldsymbol{u}(\boldsymbol{x}, t)$ (a Brownian noise should be included for the case with diffusion). The Jacobian $J = \det \mathbb{J}$ is the *de*terminant of the evolution matrix $\mathbb{J} = \partial \boldsymbol{x}/\partial \boldsymbol{\zeta}$, with $J^{-1}dJ/dt = \nabla \cdot \boldsymbol{u}$ and $J(0) = 1$. Then

$$\theta[\boldsymbol{X}(t; \boldsymbol{\zeta}, 0), t] = J^{-1}\theta[\boldsymbol{X}(0; \boldsymbol{x}, t), 0], \tag{2.8}$$

$$c[\boldsymbol{X}(t; \boldsymbol{\zeta}, 0), t] = c[\boldsymbol{X}(0; \boldsymbol{x}, t), 0]. \tag{2.9}$$

In words, it is simply the fact that, along the trajectory, the density $\theta$ is compensated by the effect of volume change and the tracer $c$ stays constant (thus the value of the initial time). For the density, we have the classical result

$$-\theta^{-1}d\theta/dt = \nabla \cdot \boldsymbol{u} = J^{-1}dJ/dt \tag{2.10}$$

from Eq. (2.6): $\theta = \theta_0 \exp\{-\int_0^t \nabla \cdot \boldsymbol{u}(\boldsymbol{x}', t')dt'\}$ along the fluid particle trajectories $\boldsymbol{x}'$ starting with $\theta_0$, or $\theta$ and $J$ have locally exponential dependence on the velocity divergence whose fluctuations cause highly intermittent response. The density then becomes larger in the compressed region with negative $\nabla \cdot \boldsymbol{u}$, and vice versa, i.e., $\theta$ concentrating on smaller and smaller regions, eventually curdling on fractal clusters and leaving more and more voids; more general situations with diffusion and random forcing present of course much more complex evolutions, but such a fundamental mechanism should still be at work [44, 36, 4]. Actually, from Eq. (2.6), we have

$$\partial_t \frac{\theta^\ell}{\ell} + \nabla \cdot \frac{\theta^\ell \boldsymbol{u}}{\ell} = \frac{1-\ell}{\ell} \theta^\ell \nabla \cdot \boldsymbol{u} \tag{2.11}$$

with the real power index $\ell \neq 1$. With spatial (and statistical) averaging over Eq. (2.11), the above discussion following Eq. (2.10) indicates positive correlation between $\theta^\ell$ and $-\nabla \cdot \boldsymbol{u}$ for "high" value of $\ell$ and that high-order moments should grow: see Ref. [4] and references therein for more systematic computations with the quantitative results implying $\ell > 1$ is "high". Thus, sufficient diffusion (to prevent the indefinite growth of the fluctuation) of the density $\theta$ is in a sense crucial, which should be kept in mind in the discussions below. When the flow is incompressible, $\theta$ and $c$ share the same dynamics, which is part of the reason for us to write down the $c$-equation to emphasize the compressibility effect on $\theta$. In general compressible flows, $c$ traces the fluid particles and presents very different dynamical properties. The variance of $c$ is not dynamically invariant but is statistically, and it has been argued that a phase transition of cascade directions happens with increasing compressibility [4], the detailed discussion of which, among others, is bedyond the scope of this note.

From the above discussions, it is then intuitively clear that, even with the diffusion, larger compressibility in general leads to larger variance of $\theta$. The helicity fastening effect has then the consequence of reducing the $\theta$ fluctuations, which can be made to be more explicit by examining the energy equation of $\theta$ ($\ell = 2$ in the above), which, like the others, is just the same as the case for the inertial particles in Sec. 2.1.2 below and is deferred to Sec. 2.2 [Eq. (2.16)].

### 2.1.2. Consequences on inertial particles

Inertial particles in different situations should be modelled differently, ranging from passive and mutual-coupling ones, for distinct situations. We now consider the passive inertial particles with a number density $n_p$ and neglect the diffusion and viscosity for simplicity in the following discussion. The inertial particle velocity field $\boldsymbol{v}$ satisfies

$$\partial_t \boldsymbol{v} + \boldsymbol{v}\nabla \cdot \boldsymbol{v} = \nu_p \nabla^2 \boldsymbol{v} + \boldsymbol{f}_p, \tag{2.12}$$

where the diffusion term [17] is neglected for brevity, and the equation for $n_p$, Eq. (2.15) below with the diffusion and pumping neglected, is of the same form (2.2) of $\theta$ for non-inertial particles written before, with $\boldsymbol{u}$ replaced by $\boldsymbol{v}$. Various models for $\boldsymbol{f}_p$ have been adopted for different purposes (e.g., Refs. [46, 20, 4, 17]). The simplest and widely used model of the $\boldsymbol{f}_p$ is the friction drag

$$\boldsymbol{f}_p = (\boldsymbol{u} - \boldsymbol{v})\omega_p, \tag{2.13}$$

with $\omega_p^{-1} = \tau_p$ being the response/stopping time. $\tau_p$ can in general depend on space and time (with $\bar{\rho}$ below replaced by the local $\rho$ [18] say) but is sometimes, for analytical tractability, taken to be a constant, $\sqrt{\frac{\pi}{8}}\frac{\rho_p a}{\bar{\rho}c_s}$, depending on the particle grain size/radius $a$, material density $\rho_p$, and, the carrier flow mean density $\bar{\rho}$ and sound speed $c_s$: see, e.g., Ref. [26], where the additional potential, thus nonhelical, gravity force (for the acceleration and clustering of cosmic dust), is also included without affecting the discussion below. The model (2.13) is a quite universal component of more complicated models, which is important to the generality of the helicity transfer to be argued below.

Note that $\tau_p$ somehow characterized the slaving degree of $v$ to $u$, since $v = u$ solves the system when there is neither forcing nor dissipation (or dispersion) differences. In the literature, another time scale $\tau_u$ of $u$ is introduce to define the Stokes number $\tau_p/\tau_u$ for characterizing the physics. We note that $\tau_u$ can be chosen for large-scale (depending of the integral scale, say) or small-scale regimes (depending on the dissipation scale), among others, and another time $\tau_v$ can also be introduced, nonuniquely again, due to the fact that the $\tau_p$ term is for the coupling of the $u$-$v$ system. So, we do not bother to make such an effort on introducing the Stokes number here. It is sufficient for us to recognize its coupling or slaving role.

The equation for $\boldsymbol{v}$ is nothing but Burgers', up to a linear dragging term $-\omega_p \boldsymbol{v}$: this is highly nontrivial, because, presumably, the helicity fastening effect in the particle flow then can be large according to our previous discussion (Fig. 12 below), even with $\boldsymbol{u}$ of the carrier flow being incompressible. So, the helicity fastening effect and its consequence on particles discussed in this note is particularly relevant for particles in various realistic problems ranging from lead particles from the exhaust pipes to cloud condensation nuclei for water droplets, and to cosmic dust. The only case not as intimate in this respect is the noninertial particle in incompressible turbulence (where, as said, the density scalar $\theta$ becomes a tracer scalar like $c$ the helicity effect on which however is a different issue.)

*Transfer of helicity.* We then see that helical $\boldsymbol{u}$ can directly inject helicity into $\boldsymbol{v}$, leading to the fastening effect on the latter, thus the consequences on $n_p$ like those for $\theta$: this is obvious when $\tau_p$ is taken to be a constant. Actually, for low relative Mach numbers, $\omega_p$ in the widely used Stokes or even the Epstein drag [47] is proportional to $\rho$ (e.g., Ref. [17] and references therein). Now, since $\omega_p$ is not an explicit function of $\boldsymbol{u}$, we see that

$$h_p = \nabla \times (\omega_p \boldsymbol{u}) \cdot \omega_p \boldsymbol{u}/2 = \omega_p^2 h + \underline{\nabla \omega_p \times \boldsymbol{u} \cdot \boldsymbol{u}\omega_p/2}. \tag{2.14}$$

The second term in the above right-hand side vanishes as indicated by the slash, so the forcing on $\boldsymbol{v}$ has the same helicity density of $\boldsymbol{u}$ up to a factor $\omega_p^2 \geq 0$; that is, the helicity of the carrier flow is transferred into the particle flow. Such a helicity transfer mechanism between $\boldsymbol{u}$ and $\boldsymbol{v}$ should hold more generally also for other reasonable models (even with $\omega_p$ depending on $\boldsymbol{u}$), because the other terms could contribute no helicity for lack of control on the latter. Indeed, as we will further remark in the discussion of the (a)KdV(B) problem, the dominant balance of $\boldsymbol{f}_p$ implies the major helicity at energy-containing large scales should be equalized between $\boldsymbol{u}$ and $\boldsymbol{v}$.

*Transfer of (reduced) compressibility.* Since helical carrier turbulence shall have less compressibility of $\boldsymbol{u}$, which then, by Eq. (2.13), indicates that $\boldsymbol{v}$ may be less compressively forced, and thus less compressive, so the helicity fastening effect is supposed to be transferred to $\boldsymbol{v}$ through the (less) compressive component of $\boldsymbol{u}$ in $\boldsymbol{f}_p$: for constant $\omega_p$, this is obvious from $\nabla \cdot (\omega_p \boldsymbol{u}) = \omega_p \nabla \cdot \boldsymbol{u} + \boldsymbol{u} \cdot \nabla \omega_p$; in general, the second term of the right-hand side of the latter, the only possible compressive component of the dragging when $\nabla \cdot \boldsymbol{u} = 0$, is not directly controlled, but, statistically speaking, it should have not much, if any, net difference without explicit influence on the correlation between $\boldsymbol{u}$ and $\omega_p$. Such direct transfer of reduced compressibility is also supported by the dominant balance of $\boldsymbol{f}_p$ mentioned in the last paragraph. This is relevant to studies on the effects of the compressibility of the carrier turbulence on the dust (e.g., Refs. [21, 23, 24]).

*Charged particles.* We have argued earlier [8] that the helicity fastening effect also works in plasma flows, such as the magnetohydrodynamics (MHD) with magnetic helicity and other helicities in more complete models (two-fluid, say): it is natural to expect the extension to magneto-Burgers flows, which turns out to be indeed the case, with even much stronger fastening effect of the magnetic helicity [34]. So, it is also natural to expect similar consequence on the charged particles. Indeed, as an example, we may consider a reasonable model in ideal MHD with an additional (Lorentz-force) term $\propto (\boldsymbol{u} - \boldsymbol{v}) \times \boldsymbol{b}$ in Eq. (2.13) for the force on the charged particles due to the presence of the magnetic field $\boldsymbol{b}$ [18]. If we consider forced MHD with injection of only magnetic helicity (but neither kinetic nor cross-helicity), the particle flow shares the same $\boldsymbol{b}$, thus the helicity. Also, we know that the magnetic and kinetic helicities can be transformed into each other in MHD flows by the Alfvén effect (c.f., e.g., Ref. [48]), then it is expected that the particle flow can also receive helicity by such a mechanism. The injected particles from supernova ejection and stellar winds, say, may also bring helicities (see, e.g., Ref. [49] for a discussion of the solar wind chirality issue using the extended MHD to account for also the sub-ion-scale turbulence.)

### 2.2. Implications on the density and tracer scalars transported by the inertial particle flow

Among the various physical implications, we point out that a consequence of the helicity fastening effect would be to attenuate the clustering and coagulation of the dust, and all that (various derivatives, such as extinction and galactic dust evolution time scales, associated to astrophysics concerning the interstellar and even intra-galactic mediums). Note that we are now referring to the helicity fastening effect on $\boldsymbol{v}$ which transports the particle number density $n_p$,

$$\partial_t n_p + \nabla \cdot (n_p \boldsymbol{v}) = 0 \qquad (2.15)$$

(with the diffusion [17] and pumping, from supernova explosion, say, neglected for brevity).

The previous discussions concerning $\theta$ and $\boldsymbol{u}$ associated to the consequence of helicity fastening effect carry over. Note that now the consequence on the inertial particles also applies to incompressible $\boldsymbol{u}$; that is, the helicity of the incompressible turbulence also reduces the compressibility of the inertial particle velocity laden therein, thus relevant to objects such as the protoplanetary discs where compressibility is relatively weak, if not incompressible [25].

And, the consequence of helicity fastening effect on weakening the inertial particle clustering and coalescence can be two-fold: one is due to the transfer of helicity from $\boldsymbol{f}_p$ (following Paragraph "Transfer of helicity"), through $\boldsymbol{u}$, to $\boldsymbol{v}$ to reduce the compressibility of the latter (thus the clustering of inertial particles); and, the other, partly based on a recent result of Ref. [24] about the coagulation dependence on the carrier turbulence compressibility, is due to the reduction of the compressibility of $\boldsymbol{u}$ (with helicity) and transferred to the particle flow through $\boldsymbol{f}_p$ (following Paragraph "Transfer of (reduced) compressibility"), and thus that of $\boldsymbol{v}$.

From Eq. (2.15) with the diffusion (parameterized by $\kappa_p$) included, we have

$$\partial_t \frac{n_p^2}{2} + \nabla \cdot \frac{\boldsymbol{v} n_p^2}{2} = -\frac{n_p^2}{2} \nabla \cdot \boldsymbol{v} + \kappa_p [\nabla^2 n_p^2 / 2 - (\nabla n_p)^2] \qquad (2.16)$$

the spatial average (denoted by the overline) of which becomes, under appropriate (say, periodic) boundary conditions with no flux,

$$\partial_t \frac{\overline{n_p^2}}{2} = -\frac{\overline{n_p^2 \nabla \cdot \boldsymbol{v}}}{2} - \kappa_p \overline{(\nabla n_p)^2}. \qquad (2.17)$$

Further statistical average (over the pumping or over time with ergodic assumption, say) can be taken, and is assumed (but not explicitly denoted, for brevity) when necessary. At this point, there is no difference between the discussions of inertial and noninertial particles, and, following the discussions in Sec. 2.1.1) for the latter, we expect $n_p^2$ and $-\nabla \cdot \boldsymbol{v}$ be not only positively correlated, i.e., $-\overline{n_p^2 \nabla \cdot \boldsymbol{v}} > 0$, but also increase with stronger compressibility. The amplitude of the diffusion term depends on the energy level of $d$ fluctuations: one can think of their spectra related by a factor of $k^2$. So, with the helicity fastening effect to have weaker/less compressibility the right hand side will balance at the lower-level $n_p$-energy. The pumping, with the assumption of effectively equal power rate (which can be realized by appropriate normalizations of the quantities for comparison), can be included, without affecting the reasoning. That is, statistically speaking, helicity reduces the fluctuating strength of the particle density in a turbulent flow.

We emphasize that, somewhat like the dispersive regularization of the Burgers equation, the compressibility reduction with helicity is not something of small correction but can be a marked change (Fig. 12 below, and even more so in the magneto-Burgers cases [34]), so the above physical consequences on the particle dynamics can be remarkable. One particular way to show the remarkable consequence is to introduce a tracer scalar $c_p$,

$$\partial_t c_p + \boldsymbol{v} \cdot \nabla c_p = 0, \tag{2.18}$$

with the possible diffusion and pumping negelected for brevity again. [$c_p$ here still can have the interpretation of the mass fraction of the composition of the mixture of the particle fluids, just like $c$ in the carrier flow discussed earlier, because the response time, $\sqrt{\frac{\pi}{8}} \frac{\rho_p a}{\rho c_s}$ as mentioned earlier, for the force model in Eq. (2.13) can be the same with the same $\rho_p a$, thus treated as a one-flow model, for distinct particle species of material density $\rho_p$ (and radius $a$).] From the analyses of high-dimensional Kraichnan model and one-dimensional Burgers turbulence (c.f., Ref. [4]), it is favorable to assume the possibility of the inverse cascade of the passive scalar energy in three-dimensional Burgulence with enough compressibility. Since $\boldsymbol{v}$ is governed by the pressureless equation, its compressibility and the helicity fastening effect should respectively follow essentially those of the Burgers flows remarked earlier. So, as for $c$ transported by Burgers $\boldsymbol{u}$, $c_p$ now may also present the phase transition in its energy cascade direction characterized by different amounts of helicity from $\boldsymbol{u}$, injected by $\boldsymbol{f}$, which can be checked numerically: note that this can happen even in the incompressible-$\boldsymbol{u}$ carrier flow.

### 2.3. The dispersion effect of (a)KdV(B)

There are multiple reasons to connect the above Burgulence helicity fastening effect and consequences (HFEC) with the dispersion regularization effect and consequences in KdV(B). Beyond the analogy, it is also an attempt to understand and model the mechanism of HFEC, by establishing relevant ideas in three-dimensional (3D) space, with particularly the construction of three-space structures that would correspond to those (presumably helical) responsible for the HFEC in turbulence. We now slightly elaborate them.

#### 2.3.1. Helicity regularization and dispersion regularization effects

The dispersion regularization effect of KdV associated to the introduction of waves is old and familiar (in the Eulerian description), without the necessity of elaboration (but see Sec. 2.3.2 below from the Lagrangian point of view which, to the best of our knowledge, has not been taken for this issue). The notion that the helicity fastening effect leads to regularization (at least to some degree) of the three-dimensional Burgers flow appears also clear in terms of weakening or reducing the compression that otherwise would result in more/stronger shock(s). Analogy or model of the helicity fastening effect with the dispersion instead of diffusion may be justified by several physical arguments, a particular one being that it introduces conservative oscillations and does not damp energy.

#### 2.3.2. Escaping from traps: helical motion and drifting in dispersive waves

It is well-known that the Stokes drifts present in various waves, including Rossby waves and acoustic waves (e.g., Ref. [50] and references therein), and the direct integration of the particle orbits of the traditional KdV solutions indeed results in drifts [51]. So, the regularization of the shock by the dispersive oscillations may also be understood in the Lagrangian scenario, with the particles subject to additional drifts away, countering, to some degree, the compression clustering effect, from the highly compressive location where they otherwise would aggregate upon collision [4] (trapped at the shocks). It is intuitively clear that particles can escape from the helical structures. [We will not use the Lagrangian technique to track particles in this note, but see Ref. [17] for the pros and cons of the Lagrangian and Eulerian methods.]

## 3. Dispersion effects on the particle transports: (a)KdV(B) cases

In this section 3, we analyze, with the comparison of the numerical results, the particle transports in one-dimensional (1D) Burgers, KdV, aKdV and KdVB equations all of which are models starting from the basic self-advecting inviscid-Burgers or Hopf (Burgers-Hopf) equation $\partial_t u + u \partial_x u = 0$ with different supplementations of dissipation and/or dispersion terms. Some of the numerical results for comparison being kind of elementary or pedagogical though, they are very helpful for emphasizing the dispersion effects and the differences in the consequences between KdV and aKdV dispersions with sometimes contrasting behaviors.

The KdV equation for real $u$ replaces the diffusive term in 1D Burgers with a dispersion and reads

$$\partial_t u + u \partial_x u + \mu \partial_x^3 u = 0, \tag{3.1}$$

or, with $2\pi$ periodicity, in Fourier $k$-space

$$(\partial_t - \mu \hat{i} k^3)\hat{u}_k + \hat{i} \sum_{p+q=k} q \hat{u}_p \hat{u}_q = 0, \tag{3.2}$$

where the Fourier coefficient

$$\hat{u}_k(t) = \int_0^{2\pi} u(x,t) \exp\{-\hat{i}kx\} dx/(2\pi) =: \mathscr{F}\{u\}(k,t) \tag{3.3}$$

with $\hat{i}^2 = -1$ and the complex conjugacy $\hat{u}_k^* = \hat{u}_{-k}$, thus

$$u(x,t) = \sum_k \hat{u}_k \exp\{\hat{i}kx\} =: \mathscr{F}^{-1}\{\hat{u}_k\}(x,t). \tag{3.4}$$

The aKdV equation is given by [1]

$$\partial_t u + u \partial_x u + \mu \partial_x^3 \Big[ \mathscr{F}^{-1}\big\{ \mod (k+1,2)\mathscr{F}\{u\}(k) \big\} - \mathscr{F}^{-1}\big\{ \mod (k,2)\mathscr{F}\{u\}(k) \big\} \Big] = 0 \tag{3.5}$$

with $\mod(k+1,2) = [(-1)^k + 1]/2$.

Including both diffusion and dispersion results in the (a)KdVB equation; in words, the original KdV dispersion coefficients of the alternative even- and odd-wavenumber modes are reset to be of opposite signs.

For the velocity $v$ of the inertial-particle flow, its Burgers and (a)KdV(B) versions, with the viscosity coefficient $\nu_p$ and dispersion coefficient $\mu_p$, are like the ones for $u$, except for a drag term. One purpose is to model the effect of the helicity transferred into the particle flow.

The other equations of density and tracer scalars, with diffusions, transported by $u$ and $v$ are just the one-dimensional version of those written down before. But, for clarity, we put down as an example the equation of the density $\theta$ (nondimensionalized with appropriate normalization) advected by $u$ in the logarithmic-variable ($\zeta = \ln \theta$) form which will be used in Sec. 3.1 below,

$$\partial_t \zeta + u \partial_x \zeta + \partial_x u = \kappa_\theta [\partial_x^2 \zeta + (\partial_x \zeta)^2] + \phi_\theta/e^\zeta. \tag{3.6}$$

### 3.1. Decaying, relaxation and symmetry

We study in this Sec. 3.1 the cases decaying or relaxing, without sources, from some initial fields. Starting with the initial fields of $u_0 = \sin(x)$ and $v_0 = 0$, we have the corresponding (KdV-)Burgers shock at the center of the working region, $(0, 2\pi)$, for convenience of demonstrating the (drift away from) spatial symmetry. Note that such a case corresponds to rest particles being accelerated by the carrier, and there is an roughly opposite situation where the particle is already moving and even faster than the carrier (thus decelerated), which is not demonstrated here. We focus on the acceleration case for an easy (probably simplest) way to show the transfer of dispersive oscillations and the subsequent consequences on particle transports.

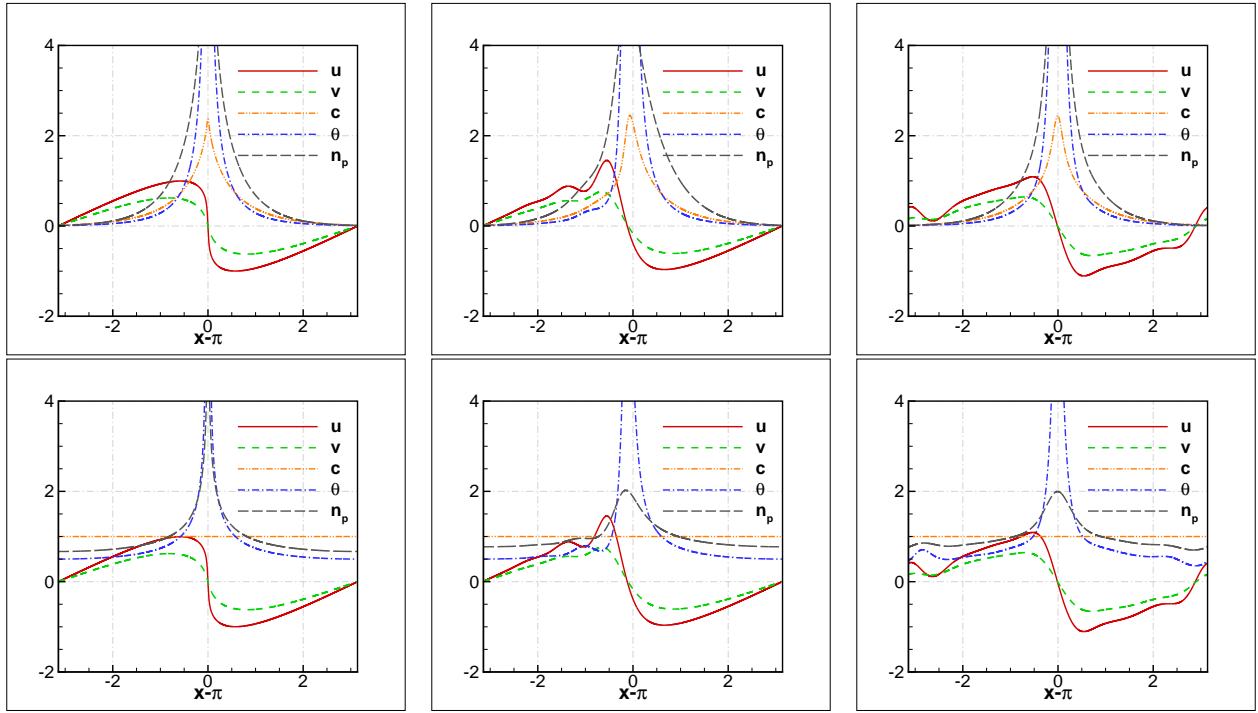

Figure 1: Burgers (left), KdV (middle) and aKdV (right) and the accordingly transported fields, starting from $u = \sin x$ and $v = 0$, at $t = 1$ decay/relax from Gaussian-like (upper) and uniform (lower) scalars.

The scenario corresponding to the phenomena, as observed by Zabusky and Kruskal [52] for the development of KdV solitons, can be described in terms of three time intervals. (I) Initially, the Burgers-Hopf terms dominate and the classical overtaking phenomenon occurs; that is, $u$ steepens in regions where it has a negative slope, and $v$ follows, with also $\theta$, $c$ and $n_p$ indicating aggregation of tracer (when initially nonuniform) and clustering of particles at the steepest location. (II) Second, the density pulse becomes narrower and stronger with the growth of the velocity negative slope and after $u$, $v$ and $\theta$ etc. has steepened sufficiently, the third term (diffusive or dispersive) becomes important and prevents the formation of a discontinuity, and different models start to behave very differently, including further emergence of more clusters following the growth of solitons from small oscillations. (III) Finally, after longer time of different developments of the carrier, some properties of particle transports turn back to be somewhat similar, in the sense that the number of clusters reduces due to collision and coalescence.

More description and analysis of the details are given below. Fig. 1 presents the snapshots at time $t = 1$, from the case with

$$\nu = \kappa_\theta = \kappa_c = \nu_p = \kappa_p = 1/400 \text{ (for Burgers)}, \ \mu = \mu_p = 1/40 \text{ [for (a)KdV] and } \omega_p = 1,$$

with two sets of initial scalars, respectively,

$$\theta_0 = c_0 = n_{p0} = 1 \text{ (upper row) and } \theta_0 = c_0 = n_{p0} = \sqrt{2\pi} \exp\{-(x-\pi)^2/2\} \text{ (lower row)}$$

(with the same total "mass" $2\pi$). It is seen that, unlike the Burgers regularization (left column) with viscosity, the dispersive regularization of KdV (middle column) with oscilations biased on one side of the shock leads to biased scalars distribution. [$c \equiv c_0 = 1$ in the lower row is due to vanishing $\nabla c$ initially, and forever.] The aKdV dispersion (right column) results in relatively very small bias (due to the differences between even and odd modes) compared to the KdV case.

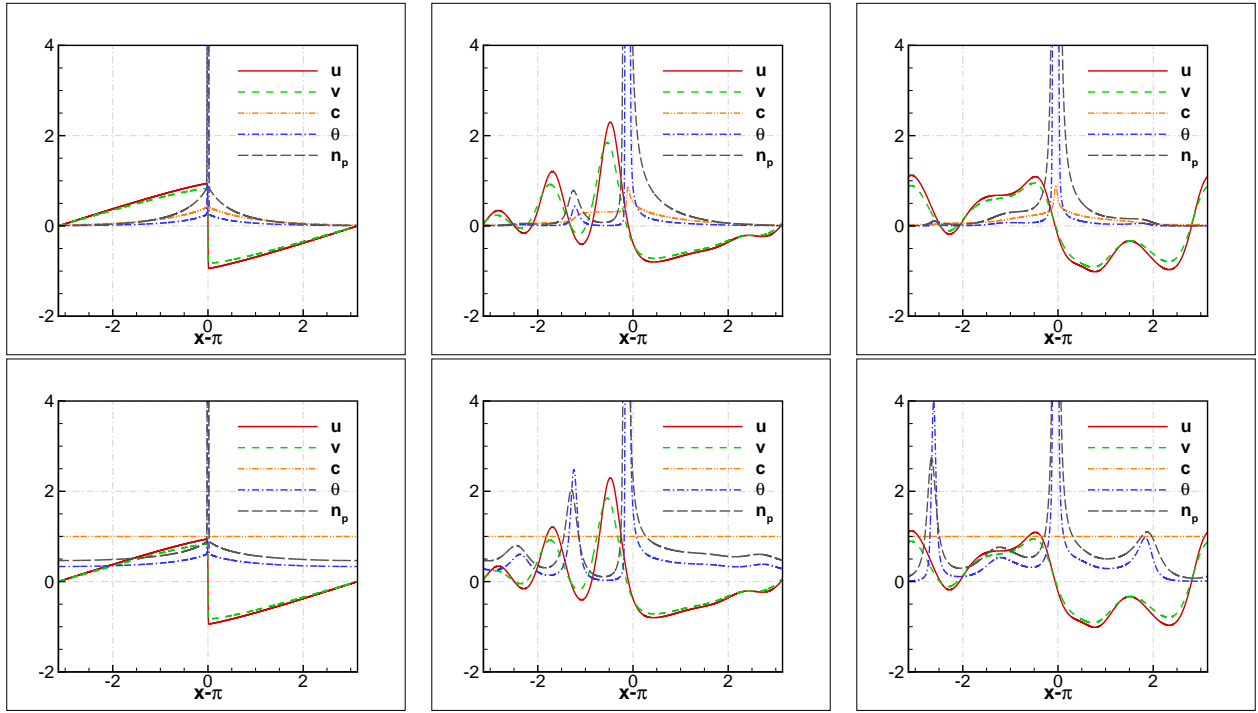

Figure 2: $t = 2$; others the same as in Fig. 1.

According to Eq. (2.10), even neglecting the diffusion effect, the Lagrangian variation of the local density is a function of the Jacobian which depends on the condition of the initial field. So, except for the shock or extremely compressed location where all fluid particles arrived are sticking together, it is not directly clear whether particles aggregate in a local region (other Eulerian local compression maxima do not necessarily correspond to the local peaks of the densities). So, it is helpful to look further into the results at later times (Figs. 2 and 3) with stronger and/or more dispersive oscillations.

$v$ eventually follows $u$ closer, and smaller $\omega_p$ (slower response with larger $\tau_p$) for "heavier" (denser and/or bigger) particles leads to more differences between $u$ and $v$, and all that (Fig. 4 with $\omega_p = 0.3$ for comparison with Fig. 2). Other modification(s) will introduce some differences. For example, $v$ may not be regularized exactly as $u$ so that $v = u$ is not the solution. Actually, if $\omega_p$ is large enough, the dragging term $f_p$ dominates when the difference between $u$ and $v$ is not too small, then dominant balance requires $v \to u$; so, incidentally, since the helicity in three-dimensional turbulence is located at large-scale energy-containing range, the helicity of $\boldsymbol{v}$ cannot be too different to that of $\boldsymbol{u}$, which is an even simpler argument, of the helicity tansferred from $u$ to $v$, than the previous one with Eq. (2.14). For example, numerical experiments (not shown) verify that, for the cases studied here but with the $v$ equation deprived of the diffusion and dispersion, i.e., with $\nu_p = 0 = \mu_p$, $\omega_p = 10$ is sufficient to have $v$ follow $u$ well enough to accordingly be regularized (without forming singularity). And, from the point of view of modeling the helicity fastening effect mentioned before, the dispersive oscillation transferred from the KdV $u$ may be enough, and, for $\omega_p$ not large enough, it can be appropriate to further regularize $v$ by the Burgers diffusive viscosity without the extra dispersive term, which results in dynamics with some flavor of KdVB (Fig. 5) which will be further discussed in the next section 3.2.

Fig. 5 also shows that the clusters of particles form at different locations of compression at the early stage and eventually merge after collisions, reducing from 4 at $t = 4$ to 2 at time $t = 8$, which indicates that

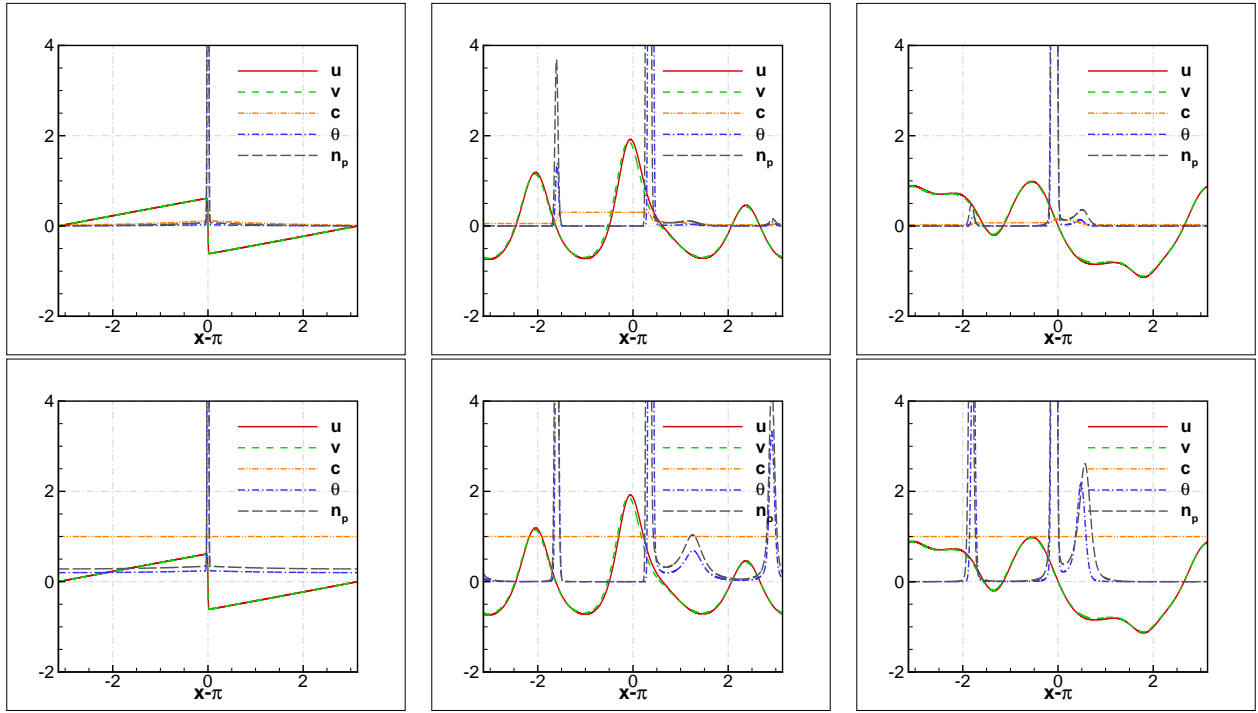

Figure 3: $t = 4$; others the same as in Fig. 1.

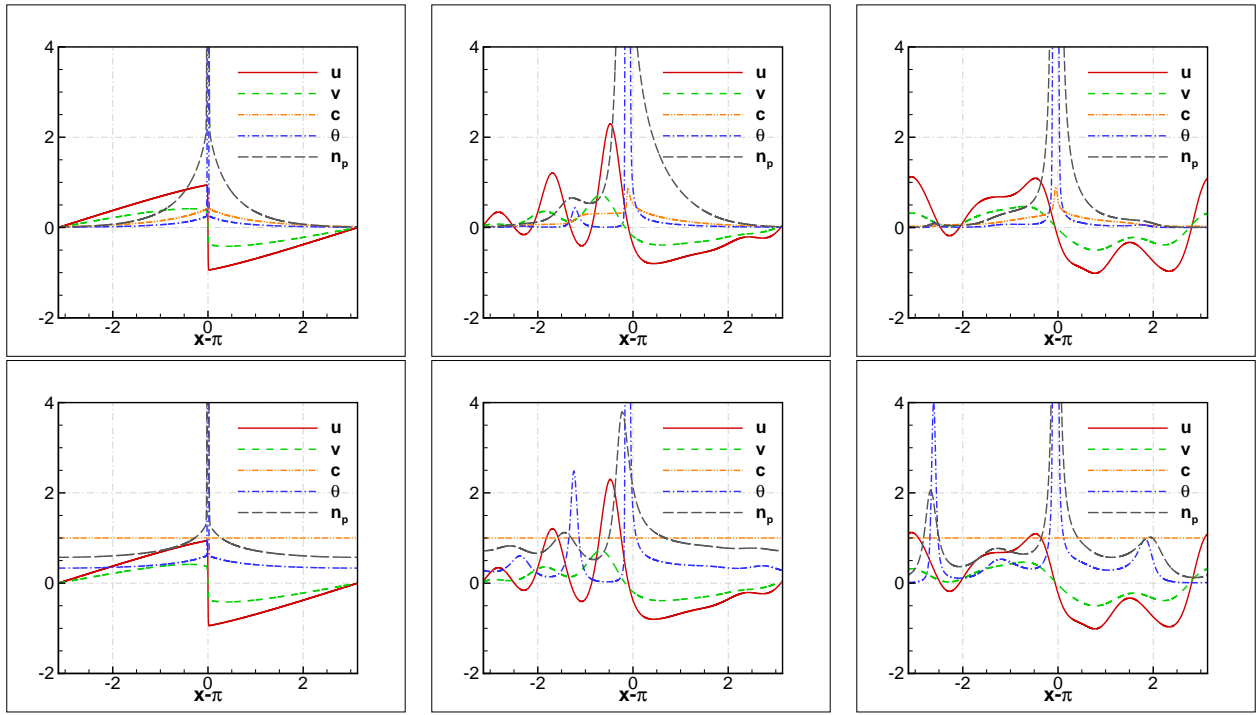

Figure 4: The case with smaller response frequency (compared to that for Fig. 2) at $t = 2$; others the same as in Fig. 1.

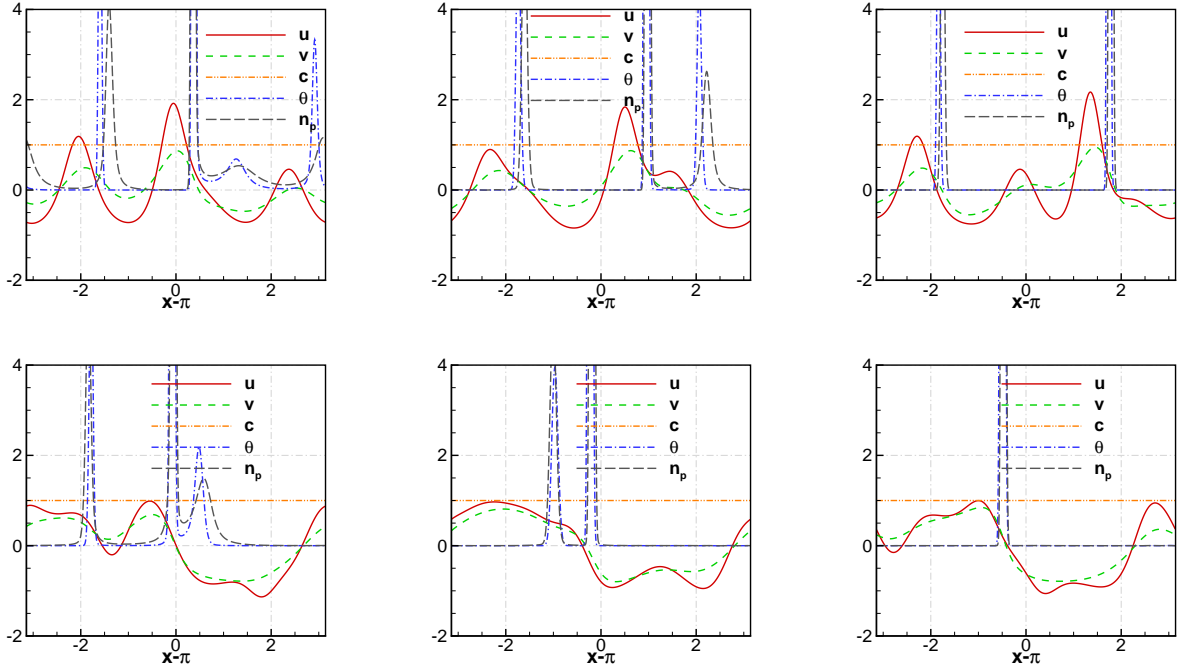

Figure 5: The KdV- (upper row) and aKdV-$u$ (lower row) cases with the dispersive term of $v$ replaced by the Burgers viscosity $0.1\partial_x^2 v$ at $t = 8$ (right), $t = 6$ (middle) and $t = 4$ (left) for comparison with, respectively, the middle and right panels of the lower row of Fig. 3; others are respectively the same as the corresponding latter ones.

to maintain many clusters distributed over space, we need randomness in the flow or from the source of the particles. As shown in I, a major difference between aKdV and KdV is that the largest (big) shock/(weak) soliton (see the discussions about "shock-soliton duality" or "shocliton" there) in the former is quite stable with very slow travelling velocity (due to the asymmetry between the wavelengths of the even-odd modes) compared to other travelling solitary oscillations. Now, the comparison of the KdV-$u$ and aKdV-$u$ cases shows faster coalescence of the clusters of the latter, i.e., all the (inertial and noninertial) particles being quickly "absorbed" by the shock(liton) in a way somewhat in between the KdV- and Burgers-carrier cases, leaving with only a single "star" travelling with the "shocliton". In this figure, the diffusion coefficients $\kappa_\theta$ and $\kappa_{n_p}$ are much smaller than the viscosity coefficients of $u$ and $v$, allowing the excitation of smaller density scales beyond the dissipation ranges of the flows, which emphasizes the fact that, if the regularization scales, dissipation or dispersion, are distinct for $u$, $v$, $c$, $\theta$ and $n_p$, richer dynamics can emerge. The main results however does not depend on such parameterization details. For example, Fig. 6 is from another simulation corresponding to Fig. 5, with the same $\nu_p = 0.1$, $\kappa_\theta = 1/400$, but, $\mu = 1/800$ and $\kappa_{n_p} = 0.1$, which allows the excitation of $u$ and $\theta$ scales smaller than those of $v$ and $n_p$ but which still presents the same scenario. We take this chance to point out that due to the formation of clusters and voids, the numerical computation of the density equations is not trivial. In the middle panel of the top row in Fig. 6, $\theta$ is particularly obviously negative and noisy around the cluster located at $x - \pi = 1.5$, which is typical in the computations with the primitive variable ($\theta$ here) for the density transition from nearly "vacuum" voids to very singular clusters and which can be avoid by using the logarithmic-variable Eq. (3.6); and similar for other scalars which are physically positive: the second row are the corresponding results from also pseudo-spectral computations of the logarithmic-variable densities, showing close results with minute quantitative differences in the regular regimes, but particularly cleaner without noisy and negative transitions. We made mutual check of such two kinds of computations for almost all cases, and have found that they are in general essentially the same as presented in this figure; and, when such (minor) numerical issue presents in the primitive-variable

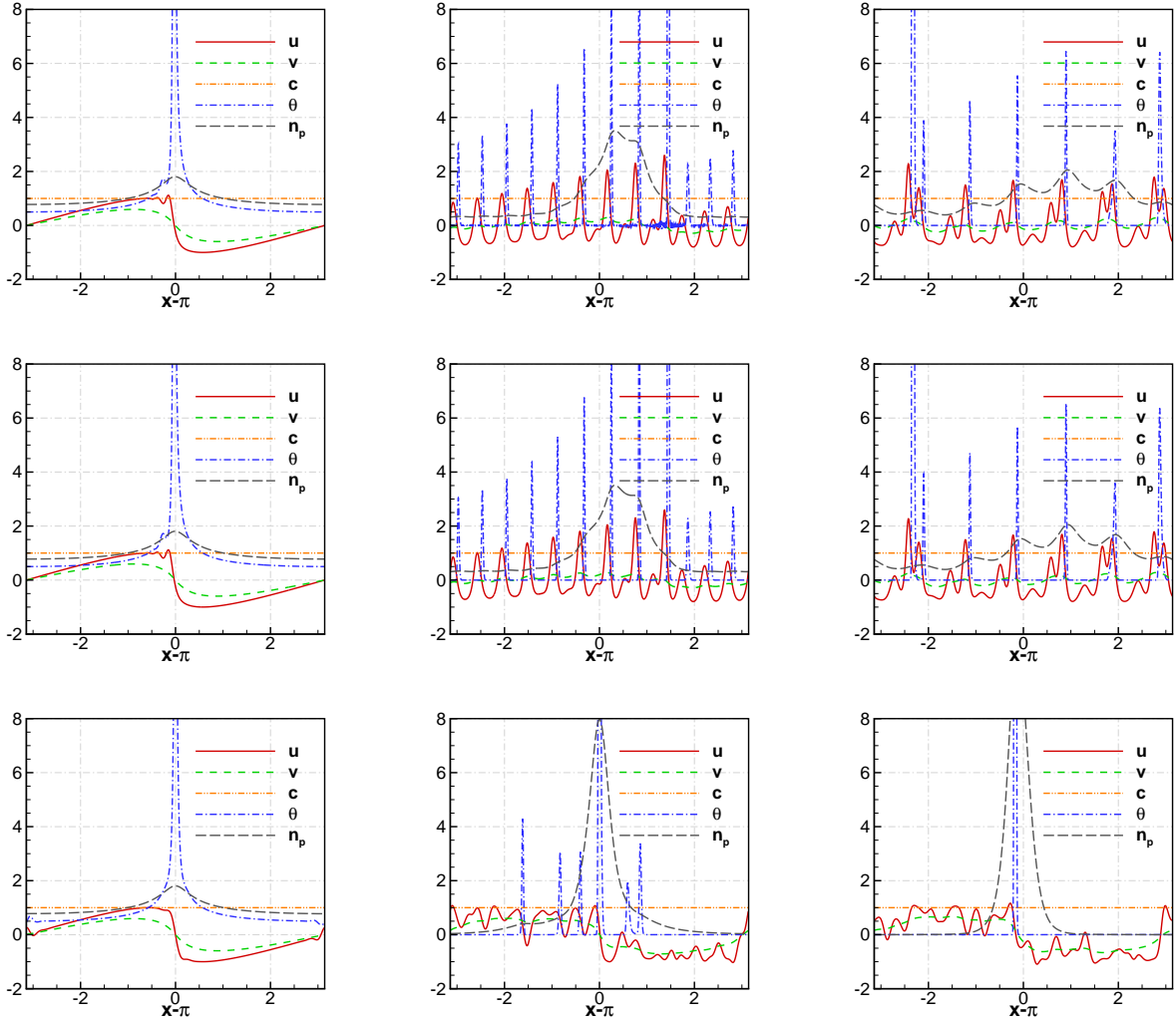

Figure 6: The KdV- (top and middle rows) and aKdV-$u$ (bottom row) cases at $t = 1$ (left column), $t = 5$ (middle column) and $t = 10$ (right column); others are the same as Fig. 5 (except for smaller $\mu$ and larger $\kappa_{n_p}$). The top row is from computing the "primitive-variable" density equations, and the middle from the logarithmic-variable ones.

computations, we use the logarithmic-variable results.

We have seen that, overall, the KdV and aKdV dispersions, compared to the Burgers, lead to milder and more local regions of particle clustering (with peaks of the density scalars $\theta$ and $n_p$ drastically different to the tracer scalar $c$). As also intuitively natural, such a conclusion pertains with some degree of genericity, although the results are from particular setups. Even though the dispersion causes lasting oscillations, the latter only supports few clusters (actually one, in the end), opposite to the fractal mass distribution as in the inviscid limit of Burgers with dense (!) shocks from the initial Brownian velocity ([53] and references therein). This should not be very surprising, because the latter case has a sufficiently "homogeneously random" initial conditions and self-similarly decaying dynamics, or simply speaking our case, with the help of dispersive oscillations though, does not have as sufficient random fluctuations.

We wind up this section by reiterating that the intrinsic "clocks" (determined by the structures and parameters) may be different for different models, so the serious quantitative comparison between the results from Burgers and (a)KdV is in general impossible. For example, under appropriate conditions in the KdVB equation with both dispersion and dissipation, $\mu \sim \nu^2$ can be taken as a criteria for distinguishing diffusion or dissipation dominated (e.g., Refs. [54, 55]). This point however does not affect our conclusion: small $\nu$ is particularly taken to clearly indicate the asymptotic shock (regularized by the minute viscosity) and to emphasize the powerful regularization effect of the dispersion.

### 3.2. The (a)KdVB with sources (forcing and pumping)

In Sec. 3.1, we have compared some very basic decay or relaxation properties of the particles transported by the Burgers, KdV(B) and aKdV(B): Those results are "minimal" in the sense that there are still many others to be explored and that they are sufficient for the purpose here, sufficiently illuminating and motivating for us to turn to the (a)KdVB with sources (forcing and pumping) in particular.

Summarizing the discussions in the last sections, it is necessary for us to further analyze the particle transportations by (a)KdVB flows with random forcing and particle injection (pumping of the density scalars in some way). For physical context, one may think of, for instance, supernova explosion and particle adhesion for, respectively, the source and damping.

We want to study the (statiscally) steady state, so the carrier flow $u$ is driven at large scales randomly and dissipated at small scales by the viscous term, with the dispersive term also included (to model the helicity fastening effect at our current context); that is, a random forcing and a viscous damping are added to Eq. (3.1) or (3.2). The noninertial particles density $\theta$ can be further randomly pumped independently to the randomness of the transportation passively by $u$, and a diffusion term balance the system to maintain a statistical steady state. We can (but not necessarily) also introduce forcing for the inertial particle flow $v$ which is already dragged by the random $u$ with dispersive oscillations, and only a viscous term is already good for regularization, with an additional option of adding a dispersive term though. The inertial particle density $n_p$ and the tracer scalar $c_p$ passively transported by $v$ are also pumped independently and damped by molecular diffusion.

To be clear, we put down explicitly the KdVB carrier and particle system equations mentioned in Sec. 3:

$$\partial_t u + u\partial_x u + \mu\partial_x^3 u = \nu\partial_x^2 u + f, \tag{3.7}$$

$$\partial_t v + v\partial_x v + \mu_p\partial_x^3 v = \nu_p\partial_x^2 v + f_p, \tag{3.8}$$

$$\partial_t n_p + \partial_x(n_p v) = \kappa_p\partial_x^2 n_p + \phi_{n_p}, \tag{3.9}$$

while those for $\theta$ and $c$ are just the one-dimensional version of Eqs., respectively, (2.2) and (2.3); see Eq. (3.6) as an example for the logarithmic-variable form. The $f_p$ in Eq. (3.8) can have an independent random component besides the dragging term $(u - v)\omega_p$, and the $\mu_p$ term can be set to zero, with still (dispersive) oscillations transferred from $u$.

Note that, without source, the diffusive equation for the density conserves the total mass (density integrated over space). And, due to compressibility working somehow as a source the fluctuations of the density (measure by the variance, say), the density can reach a statistical steady state without further pumping; the latter of course can be included. So, it makes sense to study both two randomly driven flow systems, with and without density pumping. For nonvanishing $\phi_\theta$, a realistic modeling should be made according to the physical situations, but a particular ansatz of $\phi_\theta = s_\theta\theta$ is appealing, with the intuitively reasonable picture that large/heavier clusters overall attract more particles, for greater gravitation or sitting at the locations of larger compression, with of course randomness represented by $s_\theta(x)$ (independent of $\theta$); similarly for $\phi_{n_p}$.

#### 3.2.1. Unpumped particle densities transported by driven flows

If the Burgers equation is only driven at large scales (low-$k$ modes), the statistical equilibrium profiles of $u$ are somewhat boring in the sense that the profiles are in general left with few shocks (in general only one). Similarly would be the (a)KdVB model. For hyperviscous Burgers with self-similar random forcing, the fields appear to present rich shocks [56], well approximating the Burgers with vanishing viscosity [57], which then led us to wonder what would be the case for Burgers with normal viscosity. If shocks are also rich, then the density could present interesting distribution. In particular, the dispersion should add more fun to the problem with oscillations. So, we choose to force the system with a power-law of exponent $\beta = -1$ for the power spectrum of the external acceleration working over the full range ("all-scale forcing").

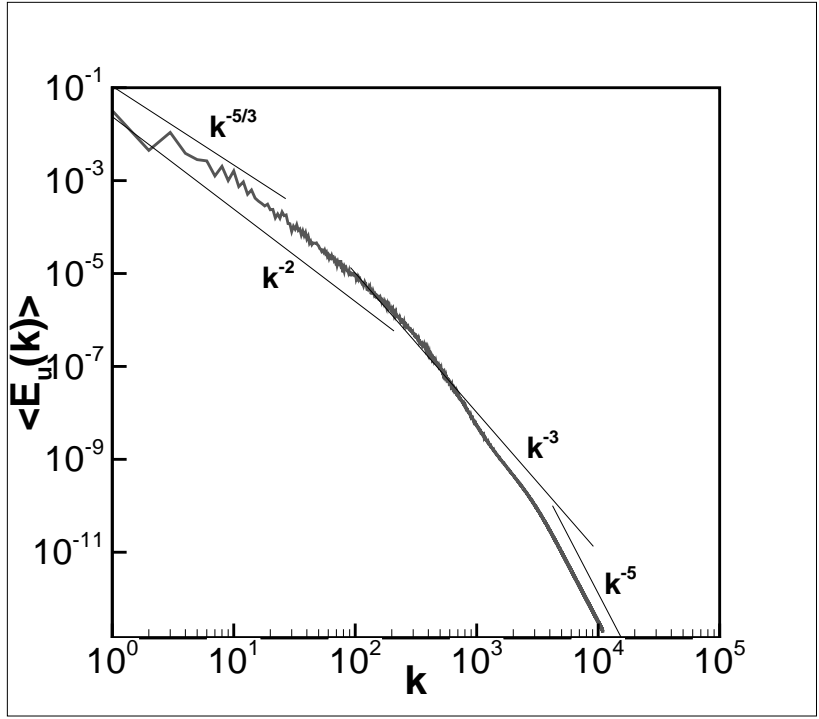

Figure 7: Time averaged energy spectrum of the Burgers $u$ with all-scale forcing. Thinner lines are added with reference scaling laws, as designated.

For the all-scale-forcing case, the (time-averaged) energy spectrum indicates several regimes of different scaling laws. There were different arguments and numerical results (see, e.g., Ref. [57] for a summary) supporting, respectively, $k^{-5/3}$ and $k^{-2}$ for the same type of Gaussian forcing $f$ with $\langle \hat{f}_k(t) \hat{f}_k'(t') \rangle \propto k^{-1} \delta_{k,-k'} \delta(t-t')$. Our numerical result (Fig. 7 for the power/energy spectrum $E_u = \sum_k |\hat{u}|^2/2$, and similarly defined for other spectra to be discussed below, with time average $\langle \cdot \rangle$) seems to indicate that there may actually be subsequent regimes dominated, respectively, by $k^{-5/3}$ and $k^{-2}$ scaling laws, which can be due to the distinct balance or dominance of different dynamical ingredients for larger and smaller scales. Which one will present or at what a scale the transition should happen depends on the setups that determine the relative strength of various elements. Similarly is in the even smaller scales where dissipative damping is effective and where the spectrum is designated subsequently with two reference scaling laws, respectively, $k^{-3}$ (c.f., Ref. [58]) and $k^{-5}$. The last regime contains little energy and is not really needed to be resolved for usual purpose, without influencing the accuracy. We are not supposed to particularly discuss such subtleties but just want to take the chance as a warmup to note the possibility of different structures dominating at different scales and, for a temporary digression in passing to emphasize that, as we shall see (Sec. 4), the (a)KdV dispersive oscillating structures may conveniently provide players to form new scalings different to the shock-dominated $k^{-2}$ and to the dissipation-dominated exponential decay.

Fig. 8 presents the profiles of all computed variables in a forced Burgers and the transported noninertial- and inertial-particle system, including the two tracer scalars $c$ and $c_p$ mentioned before, at three typical times corresponding respectively to developing (far-from-equilibrium), near- or pre-equilibrium and equilibrium stages. The computations were carried out with $N = 2^{15}$ modes, including the non-dealiased ones, $\nu = 1/800 = \nu_p = \kappa_\theta = \kappa_c = \kappa_{c_p}$ ($\mu = \mu_p = 0$) and the forcing scaling discussed in the last paragraph (whose Fig. 7 is from this case) but $\hat{f}_k = A_u k^{-1/2} \exp\{\hat{i}\alpha_u\}$ with $A_u = 10$ and the random $\alpha_u$ uniformly distributed in $[0, 2\pi]$, and, similarly $A_c = 5 = A_{c_p}$ and independent $\alpha_c$ and $\alpha_{c_p}$, with the same uniform distribution, but $\hat{\phi}_{c;c_p} = A_{c;c_p} \exp\{i\alpha_{c;c_p}\}$ for only $|k| \leq 3$ (otherwise vanishing).

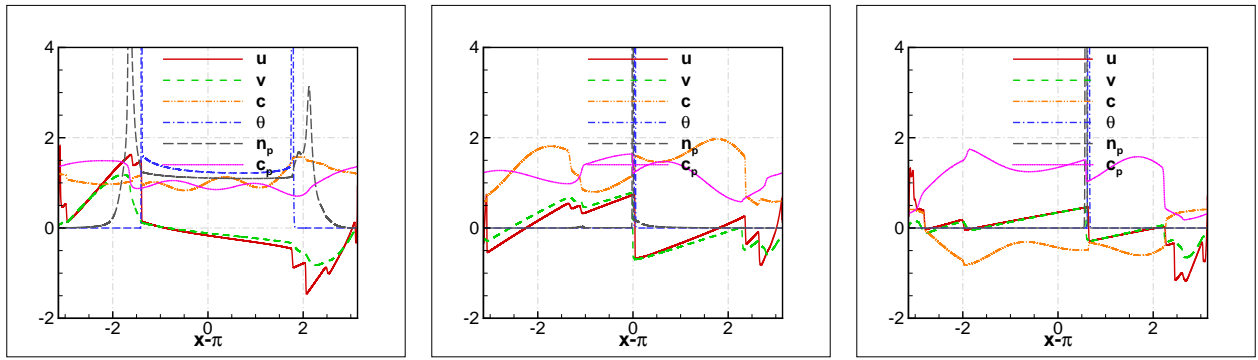

Figure 8: The Burgers-$u$ case at $t = 2$ (left column), $t = 10$ (middle column) and $t = 20$ (right column).

There are not many small-scale structures inbetween shocks, unlike the hyperviscous [56] and inviscid-limit [59] cases. This is due to our application of normal viscosity and finite dissipation scale, not because of the uniform phase different to the Gaussian $\hat{f}_k$ (we have reproduced essentially the same structures — not shown — of Ref. [56] with the same hyperviscosity but with our forcing scheme). We tend to believe that for any finite $\nu$ (and $\kappa_\theta$), the same scenario with few clusters (for both noninertial and inertial particles) located at the major shocks as presented in Fig. 8 should hold. Furthermore, the same scenario in the inviscid limit of Burgers for $u$ seems to be supported by the fact that still only few major shocks present in the inviscid limit (unlike the Brownian initial field problem with no forcing [53]), which however deserves more rigorous analysis. [Note that, with finite viscosity in KdVB, the vanishing dispersion limit is regular, not as in KdV; in other words, the limit of $\mu \to 0$ and then $\nu \to 0$ is trivially that inviscid limit of Burgers.] Our concern is whether the dispersion oscillations in (a)KdVB can change the scenario.

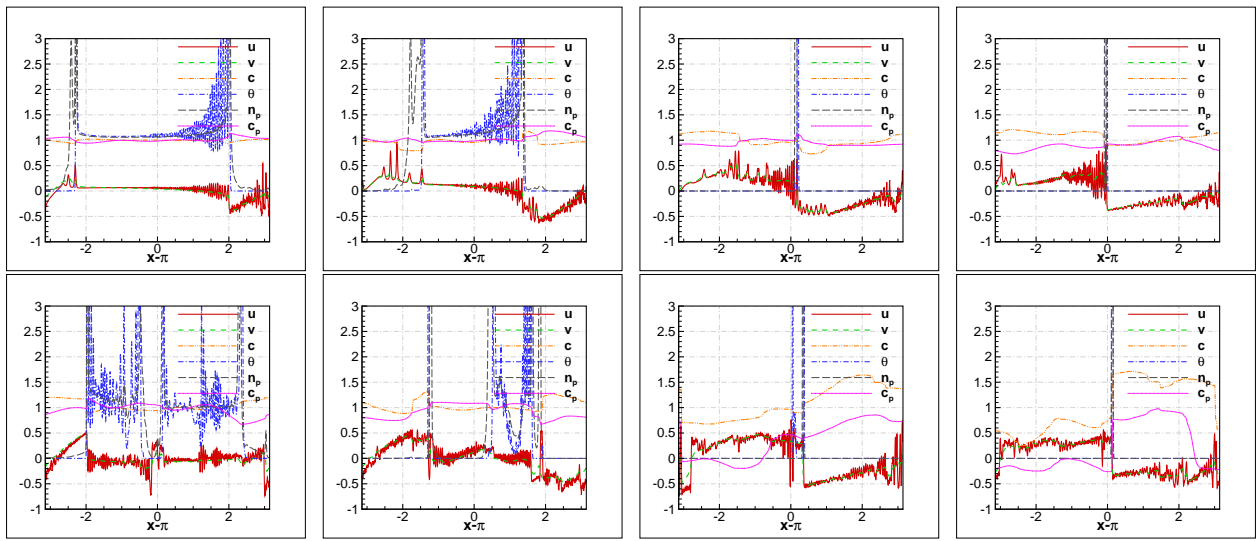

Figure 9: The KdV- (upper row) and aKdV-$u$ (lower row) cases at $t = 4$, 8, 20, 24 (subsequently the first and the fourth column from left to right): $\omega_p = 1$; $\nu = 2.5/40000$, $\mu = 20/1000000$; $\nu_p = 50/40000$, $\mu_p = 0$; $\kappa_\theta = \kappa_c = \kappa_{n_p} = \nu_p$; $A_c = 2A_{c_p} = 2 = A_u$; others the same as for Fig. 8.

The limit, if exists, with $\nu \to 0$ first and then $\mu \to 0$ is intriguing. If we take a hyperviscous Burgers with large hyperdissipativity (which is 6 in Ref. [56]) as a good approximation of the inviscid limit (with entropy solution), then adding decreasing KdV dispersion to it seems to be able to offer some information for such a limit. The dispersion coefficient however should not be too small, otherwise we would already fall in the limit of hyperviscous Burgers, i.e., turning into the "trivial" limit of $\mu \to 0$ first and then vanishing (hyper)viscosity: we are not able to discuss such a highly nontrivial and subtle issue in this note, but eventually studying the same issue of aKdVB may be mutually helpful; so, here we first provide some relevant comparisons, as presented in Fig. 9 which indeed shows that the aKdVB cases have richer (in the sense of more homogeneously distributed oscillations) structures with the densities following them, but finally all clusters coalescing into a single one located at the major shock. It takes longer for the clustering and coagulation of the particles in aKdVB with the same parameters as in KdVB. [The tracer scalars are again insensitive to the velocity fluctuations. Note that negative values can appear with our random pumping which can be negative, and we did not use logarithmic-variable computations.]

The above results indicate that shock should be excluded to have well distributed particles. This is possible in the Burgers case, by raising the exponent $\beta$ to be positive [60, 61]. However, with the dispersion (our interest in this communication), shocks still present and condensate nearly all particles, as confirmed by our computations (not shown); so, we do not further pursue such a study here.

### 3.2.2. Pumped densities for particles (with injection) transported by driven flows

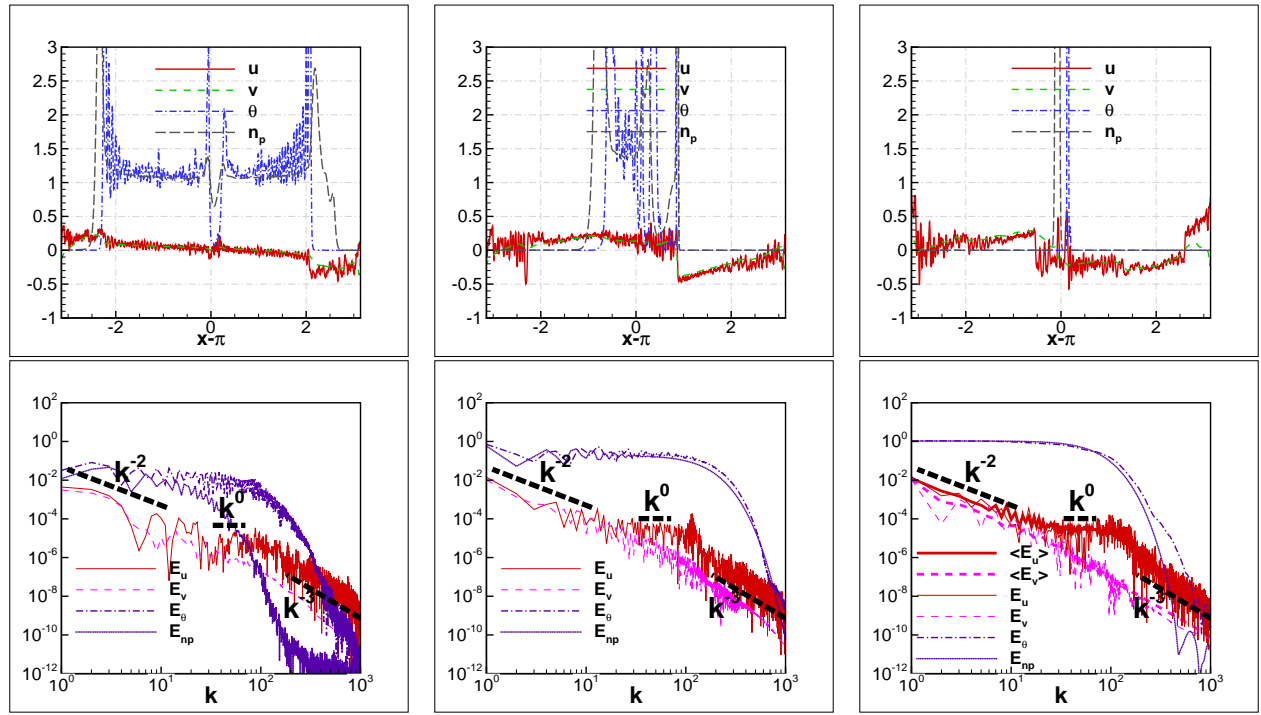

Figure 10: The profiles (upper row) and power spectra (lower row) at $t = 4$, 12, 76 (subsequently from left to right) for pumped particle densities with driven aKdV-$u$ and Burgers-$v$: $\omega_p = 0.3$; others same as for Fig. 9. On the right panel of the lower row, the energy spectra of $u$ and $v$ averaged (over time) around the final time $t = 76$ in the statistically steady state are also plotted, while the snapshots of $\theta$ and $n_p$ at that moment are already quite smooth and very close those averaged ones which are thus not shown. Much thicker long-dashed lines are respectively for three reference scaling laws ($k^{-2}$, $k^0$ and $k^{-3}$) at the appropriate scale regimes for the flow velocities.

We rather turn to the case with particle injection, modeled by a (random) term proportional to the local density as mentioned before. We let $s_\theta$ and $s_{n_p}$ in, respectively, $\phi_\theta = s_\theta \theta$ and $\phi_{n_p} = s_{n_p} n_p$ both but independently be uniformly distributed over $(0.75, 1.25)$. Such a model is also meant to maintain, with some randomness though, the local clusters which emerge in the early stage of the development. $v$ is also additionally driven with the same ansatz (used for the computation in Fig. 8) of forcing as but independently of $u$ ($A_u = 2A_v$), but no dispersion term is added to the $v$ equation. $\omega_p$ is now set to be the smaller 0.3. We see from the profiles in Fig. 10 from such aKdVB computations that, still, both inertial and noninertial particles finally fall into the respective single cluster (distributed clusters formed early not being maintain by the particle injection), with richer and longer-lasting structures during the process though. We present only the result of aKdVB, because the KdVB case is even worse, in the sense of poorer structures and coalescing into a single cluster faster.

For the power spectra in the lower row, the shock corresponds to a $k^{-2}$ scaling law in the energy spectrum, which is the case for both $u$ and $v$. The $k^{-3}$ scaling corresponds to what we have already remarked for the Burgers with all-scale forcing scheme, but it is interesting to see that it appears to be also followed by the aKdVB (and KdVB — not shown): note that the KdV velocity spectrum was shown in I to present exponential decay at large-$k$, like the dissipative Burgers.

Before explaining the $k^0$ scaling law for the flat intermediate regime of $u$ (Sec. 4 below), let us turn first to the power spectra $E_\theta := |\hat\theta_k|^2/2$ and $E_{n_p} := |\hat n_p|^2/2$ which obviously present the same $k^0$ scaling before their large-$k$ decaying dissipation ranges. As particularly obvious in the right panel in Fig. 10 for the profiles at late times, the densities are very narrow pulses, asymptotically the Dirac delta functions. The Fourier transform of the Dirac delta function is a constant, which then explains the $k^0$ scaling law. At earlier time, the well-separated pulses of the densities already form, thus also the asymptotically plat spectra, not as clean though.

## 4. The reward from the particle transport study: understanding the dispersive oscillations

The explanation for the $k^0$-scaling of the density spectra in Sec. 3.2.2 leads us to understand the same scaling, but in the intermediate regime, of $E_u$ in Fig. 10 similarly. Actually, the oscillations in $u$ correspond to the solitons in (a)KdV (without the dissipation term) which are also pulses which become narrower with smaller but finite dispersion, approaching the Dirac deltas of different amplitudes. So, we are led to the explanation of such equipartition of $u$-energy, with a $k^0$-scaling in the intermediate regime indicated there, by the asymptotically multiple Dirac deltas of the separated (atomized) narrow pulses. [This somehow further justifies the nomenclature of "soliton".] To put it simple, *the soliton is the derivative of a classical shock, just as the Dirac delta is the derivative of a step function*, in the weak sense (to be mathematically sound). This turns out to be precisely correct, even beyond the scaling-law regime, for the KdV soliton and the Burgers shock, because the derivative of the tanh profile for Burgers does be the $\mathrm{sech}^2$ profile for the KdV soliton (though never pointed out before, to the best of our knowledge); however, the statement for the scaling-law relation is of quite a universal nature for more general solitons and classical shocks (see below).

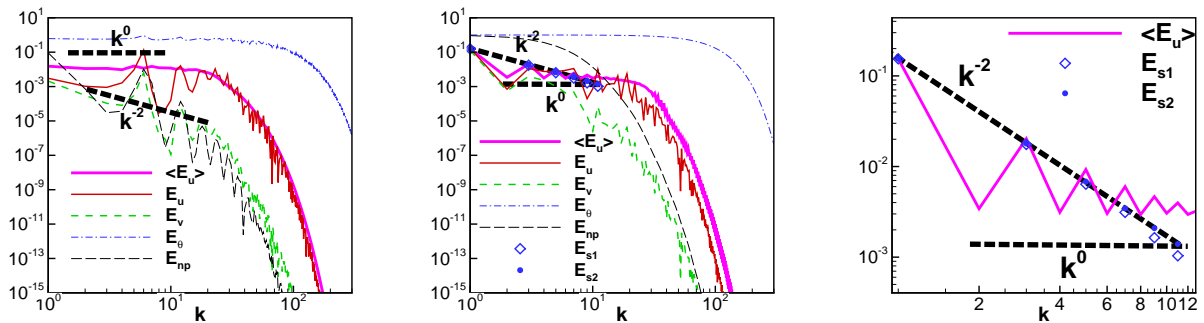

Figure 11: The power spectra of the KdV (left) and aKdV (middle, and right for the blowup) cases for some fields at $t = 10$ of Fig. 6 (the right panels there from logrithmic-variable computations), including the time averaged $E_u$. Much thicker long-dashed lines are for reference scaling laws, respectively, $k^{-2}$ and $k^0$ at the appropriate scale regimes for the flow velocities.

Indeed, we had already seen the asymptotic $k^0$-scaling in the pure dispersive (a)KdV spectra with small dispersion coefficients used in I. Such a scaling law is also seen in the appropriate late-time relaxing/decaying power spectra of densities in Sec. 3.1, and it is also observable in the (a)KdV velocity spectra there: for example, Fig. 11 for the power spectra of some fields in Fig. 6 shows that the $\theta$ spectra indeed show equipartition before decaying in the large-$k$ diffusion range but that the $n_p$ spectra however do not, for the reason that the pulses of $n_p$ are not narrow enough with too large a $\kappa_{n_p}$ here; while, both aKdV- and KdV-$u$ spectra present approximate $k^0$ scaling laws at intermediate-$k$ regimes early at $t = 10$ (but see below). Late-time results with even smaller dispersion coefficients (such as the cases in I with $\mu = \pm\mu_{ZK}/32$ for $\mu_{ZK} = 0.022^2$ as used by Zabusky and Kruskal [52]) appear to present more clearly a persistent aKdV-$u$ $k^{-2}$-spectrum at the smaller $k$s before the $k^0$-regime (followed by an exponential decay), due to the "shocliton(s)", and the KdV-$u$ (all-soliton) $k^0$-spectrum eventually occupies the whole regime, i.e., from the intermediate to the smallest $k$s before the exponential decay, like the case for the flat density spectrum, as demonstrated by the time averaged spectrum $\langle E_u \rangle$ here. [Note that the time-averaged $\langle E_u \rangle$ can be understood to be over the recursive (if indeed) snapshots and/or over the numerical errors (small here, both with an energy error $< 0.2\%$), while the statistical average issue of the solutions from a given initial field of an integrable system is subtle (see also the relevant remarks in I).]

With such a pleasant by-product reward of studying the particle transportations, we are further led to ponder more on the differences between KdV and aKdV carriers. While the aKdVB results in Fig. 10 with both diffusion and dispersion clearly present a $k^{-2}$-scaling, the latter in the averaged spectra of Fig. 11 (middle and right panels) for aKdV does not appear to be clean (neither the $k^0$-scaling in the intermediate range); the whole small-$k$ regime before the exponential decay of the aKdV-$\langle E_u \rangle$ spectrum might be observed to follow a new scaling exponent inbetween 0 and $-2$, rather than a transition from $k^{-2}$ to $k^0$ and then to the exponential decay. It then seems possible that the unique "shocliton", as a new structure introduced by the even-odd alternating dispersion, leads to a new scaling law, while the shock of aKdVB, with the viscosity turning the shock back to a classical one, still can result in a much cleaner $k^{-2}$-spectrum. Or, even further, all the aKdV oscillations (including the shocliton and anti-shocliton, in a unified sense) are fundamentally different to the KdV pulses (solitons narrowing down to Dirac deltas with decreasing dispersion coefficients) and are responsible for the new scaling behavior for the whole range before the (exponential) decay. These correspond to the problems we raised in the end of the first paragraph of the introductory discussion.

We have seriously examine the above curiousities and actually can have a finer explanation of the aKdV spectra in Fig. 11 by following closer the dynamics, with a more careful observation. As in I, the aKdV $u$-field (Fig. 6) can be decomposed into even- and odd-mode components, with the same features (thus not shown here) as in Fig. 3 of I where we see that the even-mode field $^eu$ evolves into pure pulses, like the KdV solitons, after $t_B$, and that the odd-mode field $^ou$ inherits basically the shocliton-antishocliton duo, but still with some other (solitonic) pulses. Indeed, we see from Fig. 11 (middle and right panels) that the averaged even-mode spectrum constituted by only the dip points of the $\langle E_u \rangle$ line indeed well follows the $k^0$ scaling and that the averaged odd-mode spectrum constituted by only the tip points of the $\langle E_u \rangle$ line seems to obey (roughly) the $k^{-2}$ scaling at low-$k$s, a much shorter regime than the even-mode $k^0$-range. It might appear still possible that the whole odd-mode spectrum before the exponential decay be of a different new scaling law; or, it is simply contaminated by the extra pulses following the $k^0$-scaling: the spectrum appears to be more in favor of the latter; that is, it is more likely that

$$Shocliton = Shock \oplus Soliton. \tag{4.1}$$

The superposition $\oplus$ is defined by the dynamics, and it is in general not known how the decomposition should be performed on the data, due to the possible coupling between shock(s) and soliton(s). [Here the term "soliton" is not particularly meant for the traditional rigorous soliton but more generally for the "oscillation" or "pulse" coming with the dispersion term; on the other hand, as shown in I with the contour patterns of $u$ and the odd- and even-mode components ($^ou$ and $^eu$ there for both KdV and aKdV, and other models), the latter indeed appear to follow some characteristics as the indication of traditional solitons, not very clear though.] But we still can proceed with careful observations and reasonable assumptions, turning dynamical "$\oplus$" into algebraic "+".

From the observation mentioned in the above, let's assume that the pulses of even and odd modes are statistically identical, at least at the energy level. Then, we are led to the decomposition of the energy

(spectrum),

$$E_{odd} = E_{shock} + E_{soliton} + E_{ss} \text{ while } E_{even} = E_{soliton}, \tag{4.2}$$

with the subscripts being self-evident in such a context. It is natural to expect that the coupling energy $E_{ss}$ should be relatively weak and thus negligible, when the shock is dominating low-$k$s, leading to

$$E_{shock} \approx E_{odd} - E_{even}, \tag{4.3}$$

which appears to be indeed quite accurate for odd $k \leq 11$ in the figure ("$E_{s1}$"; see below).

We naturally expect that, in $E_{even}$ at large scales, there should also be an interaction component $E'_{ss}$ which is also small but not identical to $E_{ss}$. Further observation and assumption are need to evaluate these two (higher-order) interaction terms, which appears formidable, if possible. It is possible that $E_{ss} + E'_{ss}$ resulting from Eq. (4.3) may strengthen the error in the treatment of the last paragraph, so improvement with modification is possible. For example, reasonably, $E_{ss}$ may be also equipartitioned at small-$k$s, thus can be absorbed into the equipartitioned $E_{soliton}$ to be a constant $C$. A simple linear decomposition of the energy by subtracting a same amount, i.e., the $k^0$-scaling part, from the some low-$k$ odd-mode spectrum then may work better. Besides

$$E_{s1}(2n-1) = (\langle E_u(2n-1) - E_u(2n)\rangle) * \langle E_u(1)\rangle / [\langle E_u(1) - E_u(2)\rangle] \tag{4.4}$$

corresponding to Eq. (4.3), we also plot in Fig. 11

$$E_{s2}(2n-1) = [\langle E_u(2n-1)\rangle - C] * \langle E_u(1)\rangle / [\langle E_u(1)\rangle - C] \tag{4.5}$$

for $n = 1, 2, 3, 4, 5, 6$, with $C = E_{ss} + E_{soliton}$ taken to be 0.0026 and with the respective factors $\langle E_u(1)\rangle / [\langle E_u(1) - E_u(2)\rangle]$ and $\langle E_u(1)\rangle / [\langle E_u(1)\rangle - C]$ simply for shifting the data in the log-log plot without changing the scaling property. We see in the figure ("$E_{s2}$") that the data points for $E_{s2}$ can be fitted by the $k^{-2}$-scaling (for shock) even better, more so with larger $k$s, obviously visible with bare eyes for $7 \leq k \leq 11$. $E_{even}$ can be similarly processed, but since it appears to be already quite accurately fitted by the $k^0$-scaling, we do not bother to repeat such an exercise. Such decompositions should be justified by the dynamical perturbation which is linearized when it is small at small-$k$s, which is the reason why the $k^{-2}$-scaling fits the data better at smaller $k$s.

It is natural to expect some universality of the above results among other (integrable) models, such as the Benjamin-Ono (BO) investigated in I, whose solitary pulses are also the derivatives of classical shocks, with, say, the same $k^0$-scaling of the power spectra for solitons: this has been preliminarily checked numerically, including in the data of the extended BO model computed in I with the same initial sinusoidal field (up to a $\pi/2$ phase shift) in I, with also the exponential decay at large $k$s (so, the essential difference can only be at the algebraic prefactors of the respective exponential decays.) And, the corresponding models with alternating dispersion, such as the "aeBO" model also computed in I, appear to present the same major features as aKdV we show in the above.

Note that such issues are not purely academic or minor but are on the large-scale energy-containing properties, and they should be quantitatively responsible for the different rates and strengths of the particle clustering and coalescing of the clusters shown in Secs. 3.1, 3.2.1 and 3.2.2. We hope all such efforts in this section can help eventually building a systematic theory of aKdV, as expected in I.

## 5. Further discussions

In Sec. 3, we have seen that Burgers, KdV(B) and aKdV(B) transport particles differently, all having finally the coalesced (single) cluster though. The Burgers carrier is the most "boring", in the sense that no much chance of rich structures of particle clusters. Richer multiple cluster structures can emerge and develop in the early or middel stages of the evolutions of the KdV(B) and aKdV(B) carriers. The aKdV(B) appears to be able to develop faster and richer structures and also converge to the coalesced state, with particularly the persistent shocliton, missing in the KdV(B) case, stably drifting slowly and working effectively aggregating particles and narrowing the density pulse. With finite viscosity and/or dispersion, no multi-cluster rich structures can last even with an all-scale self-similar forcing, and a pumping scheme respecting the randomness and cluster mass does not help.

An important aspect is that the dispersion (resulting in oscillations and regularization effects etc.) of the carrier can be effectively transferred to the particle flow and regularize the latter, which resembles the transfer of helicity from 3D $\boldsymbol{u}$-carrier to particle $\boldsymbol{v}$-flow and regularize the latter (to some degree). To have the comparison between the two scenarios be more concrete and more motivative, below we present some basic results of helicity fastening effect in Burgulence.

### 5.1. Helically and nonhelically driven Burgulence: Marked evidence of helicity fastening effect

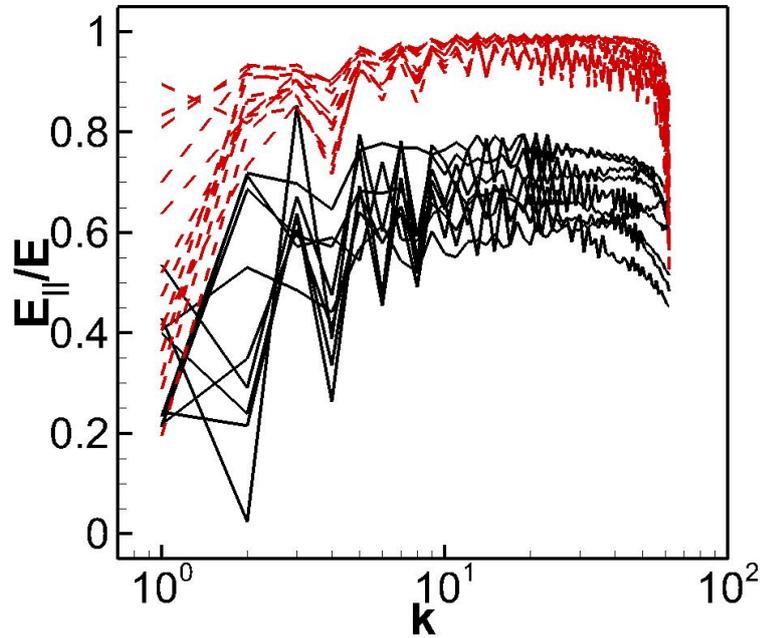

Figure 12: Snapshots from the statistical steady state of the parallel-mode spectral fraction $E_\parallel(k)/E(k)$s of helical (solid) and nonhelical (dashed) 3D isotropic Burgulence.

Fig. 12 is from a computation where the Burgulence is subject to, respectively, the nonhelical acceleration of the Taylor-Green type (parameterized by a phase $\theta$ [62]) written in the $x$-$y$-$z$ coordinate frame, and its helicalization:

$$\begin{cases} f_x^{\mathrm{nh}} = \dfrac{A}{\sqrt{3}} \sin\left(\theta + \dfrac{2\pi}{3}\right) \sin x \cos y \cos z \\ f_y^{\mathrm{nh}} = \dfrac{A}{\sqrt{3}} \sin\left(\theta - \dfrac{2\pi}{3}\right) \cos x \sin y \cos z \\ f_z^{\mathrm{nh}} = \dfrac{A}{\sqrt{3}} \sin\theta \cos x \cos y \sin z \end{cases} \tag{5.1}$$

whose maximally helicalization [63], $^R\boldsymbol{f}$, is computed from the helical decomposition of Fourier coefficients

$$^R\hat{\boldsymbol{f}} = (\hat{\boldsymbol{f}} + \hat{i}\boldsymbol{k} \times \hat{\boldsymbol{f}}/k)/\sqrt{2}, \tag{5.2}$$

of the solenoidal $\boldsymbol{f}$, representing the purely right-handed helical sector, with $\theta$ randomly chosen uniformly over $[0, 2\pi)$ at each time step. $A$ is the fixed amplitude. We define the power spectra as

$$E_{\mathrm{kin}}(k, t) = \frac{1}{2} \sum_{|\boldsymbol{k}|=k} |\hat{u}(\boldsymbol{k})|^2 \tag{5.3}$$

and the power spectrum of the compressive mode of $\boldsymbol{u}$,

$$E_{\|}(k) = \frac{1}{2} \sum_{|\boldsymbol{k}|=k} |\hat{\boldsymbol{u}}(\boldsymbol{k}) \cdot \boldsymbol{k}/k|^2, \tag{5.4}$$

thus the most detailed measure $E_{\|}/E_{kin}$ of compressibility scale by scale, as given in the figure which verifies the helicity fastening effect, with a benefit of $0.2+$ for the helical case.

Consistent similar results, from other computations with more systematic tests including the magneto-Burgulence case, together with other studies, can be found in another communication [34] in the different series for a different theme.

### 5.2. Astrophysical-chirality relevance

The reduction of compressibility with helicity has of course multidisciplinary implications and applications, including (aero)acoustics and the particular issue of particle transport in realistic situations such as astrophysics we already partly discussed in Sec. 2.

It appears to make a lot of sense talking about helical turbulence, isotropic or anisotropic, at the galaxy or galaxy cluster scale: viewing such objects as large-scale ones, we see that they are rich enough to have various situations and mechanisms for the generation of helicity; and, viewing the objects as local structures of the even larger space, we naturally expect them be helical quite often in a turbulent Universe. There are already clear observations of helical gaseous nebula in our Galaxy [64]. Actually, helicity has very fundamental origins, as said in the beginning [6, 7], which makes it ubiquitous in the Universe.

The Burgers equation for potential, thus non-vortical, velocity is widely used for the large-scale structures of the Universe [53, 57], but not completely for sure with, e.g., already other considerations of the nonlinear evolution of vortical perturbations consistently within the framework of self-gravitating motions [65] (see also the recent cold dark matter model in Ref. [11]): the precise relation of helical Burgers flow to cosmology should be left for further studies, but we expect distinct 3D shock structures than those in potential Burgers reviewed in Ref. [57].

### 5.3. Expectations

A fundamental problem concerns the mechanism of the (partial) regularization effect of helicity in the sense of reducing the strength and number of shocks. We had raised this conjecture based on some relevant analysis and later established further arguments (c.f., e.g., Refs. [8, 34]), but a systematic understanding is still lacking.

For example, part of our previous arguments [8, 66] exploit invariance laws which, *a priori*, no longer formally hold for the ideal (magneto-)Burgers dynamics. Logically speaking, it is not impossible that they hold to a sufficient degree during some important and relevant dynamical process, but it might also be that it is some other aspects, such as the chirality, in the presence of helicity that really matter. Sec. 4 offers a lesson that further detailed studies of particle transport may help understand the magic of helicity itself, and we may expect the information about the helicity conservation issue by tracking the clustering events nontrivially influenced by helicity. Another possible aspect is that, topologically speaking, helicity means knottedness which however can present without helicity [67]. Knot theory, to the best of our knowledge, so far do not have a clear answer about the essential difference between knots of different values of helicity, especially on the mechanical effects such as the fastening effect we are concerned now. Thus, much more remains to be investigated for relevant issues of (magneto-)Burgers turbulence. Now, from the analyses in Secs. 2 and 3, the dispersive model appears to be an illuminating or even directly applicable candidate:

Following Sec. 2.3.1, we recall that Ref. [8] proposed to transform the helicity fastening effect into the rotation effect for the "chiral base flow/field (CBF)" which is a locally generic helical structure (helical turbulence was viewed as a gas of CBFs). With that scenario, the Rossby waves associated to the rotation then would follow. It is then intriguing whether the CBF can accordingly be some kind of structure playing the role of solitons in "integrable" systems (c.f., the notion of soliton gas [68]) and whether such Rossby waves can correspond to the dispersive waves [oscillations in (a)KdV(B)] and be also responsible for the regularization. Thus, following then Sec. 2.3.2 with the remark that particles can escape from the helical structures such as the CBF mentioned in the above or a "tornado", much easier than from a vortex in a plane, we now have even more motivations to model the helicity fastening effect and its consequences (say, on particle transports) with the dispersion ones, besides the traditional efforts of modeling the Rossby waves with the KdV equation (and the relevant phenomena, such as the Jovian Great Red Spot, with solitons [69]).

Many waves, including the chiral gravitational waves connected with the helicity concerned here through the chiral magnetic effect (e.g., Ref. [6] and references therein), are wandering and even crowding in the Universe, which may well affect the dynamics of cosmic dust and particle cluster structures (small and large); see, e.g., Ref. [30] for the discussions on dust Alfvén waves and shocks. Solitary waves are in some sense special, but as oscillations many relevant properties of them may be shared by general waves. Our aKdV(B) oscillations provide good examples of combining shock waves and solitons as (anti)shoclitons, together with the unique particle dynamics. As mentioned in Sec. 4, the main results there are in some sense universal in various models of nonlinear waves, and the even-mode and the $shock + pulses$ decomposition of the odd-mode properties of the aKdV particularly indicate the possibility of even more general applicability in other waves. Particularly, the space-time patterns of the field component ${}^e u$ as presented in I may characterize important features of the waves in the Universe, because the latter contains various waves of distinct nature, thus travelling (almost) "independently", passing each other, following (roughly) their own characteristics. So, sound understanding and modeling of the properties of their oscillations are important and promising, as partly reflected in the seemingly digressing Sec. 4 where it was indicated that different multi-scale spectral behaviors, particularly the large-scale scaling laws, the different dynamics (rates and strengths) of particle clustering and coalescing of clusters. We expect better understanding of the astrophysical and cosmological structures, and other relevant interdisciplinary problems, from good unification of such efforts.

## Acknowledgement

The consideration of cosmic dust and particle clusters, and, their astrophysical relevance (galaxy etc.) has benefited from early interactions with Prof. Z.-H. Fan's group on cosmological structures.